\documentclass[sigplan,10pt,nonacm]{acmart}
\renewcommand\footnotetextcopyrightpermission[1]{}
\AtBeginDocument{%
  }

\usepackage{my-style}
\setcopyright{none}

\begin{document}

\title{\Large \bf Analyzing and Reducing Search Quality Differences in Vector Similarity Search}
\author{Sara Mahdizadeh Shahri, 
Martin Prammer, Jignesh M. Patel, Akshitha Sriraman 
\\ 
Carnegie Mellon University
} 
\renewcommand{\shortauthors}{}

\begin{abstract}
Modern database services scalably search over large data collections via Approximate Nearest Neighbor Search, which improves search performance at the cost of search quality, measured by \emph{recall}.
In practice, a database operator seeks to achieve a target mean recall while maximizing throughput across search queries.
We show that optimizing for mean recall masks significant differences in recall across queries even when target recall is met.
As a result, numerous queries face (1) below-target recall, hurting user experience and revenue and (2) above-target recall, wasting computation to deliver unnecessarily high search quality. 
Thus, it is critical to detect and reduce recall differences across queries.

We design \rcheck, a light-weight \runtime system that identifies low-recall queries and reduces recall differences while achieving high throughput.
\rcheck's key design principle is to dynamically, efficiently adapt search effort by increasing effort for queries below target recall and decreasing effort for those above it.
\rcheck tunes available search effort parameters, making it readily deployable. 
We evaluate \rcheck using the widely-used production-style pgvector~\cite{pgvector} database.
At the same throughput, \rcheck improves mean recall by \minRecallatIsoThroughput--\maxRecallatIsoThroughput\% and enables \minQueriesMeetingTarget--\maxQueriesMeetingTarget\% more queries to meet target recall compared to the state-of-the-art globally-tuned configuration.

\end{abstract}

\settopmatter{printfolios=true}
\maketitle
\pagestyle{plain}

\section{Introduction}
\label{sec:intro}

Modern data management services perform \textit{embedding-based similarity} search over large collections of data items, such as text, images, and videos~\cite{guha2003semantic, maoro2023leveraging, lewis2020retrieval, gao2023retrieval, acharya2023llm, wang2024towards, tian2024mmrec, huang2023dsqa, zhuang2023toolqa, jing2025large}. 
To enable such search, data items are encoded as high-dimensional \emph{embedding vectors}~\cite{johnson2019billion, pennington2014glove, mikolov2013efficient} and stored in a \emph{vector database}~\cite{pgvector,douze2024faiss,wang2021milvus,weaviate}.
These vectors form an \emph{embedding space}, where nearby vectors map to similar items~\cite{johnson2019billion,pennington2014glove,mikolov2013efficient}.

To search similar items for a query, the service encodes the query as an embedding vector and finds the closest stored vectors, i.e., nearest neighbors.
In principle, we can find nearest neighbors by comparing the query vector against all stored vectors in the database. 
But, such exhaustive search is computationally expensive at modern scales~\cite{ren2020hm, johnson2019billion}. 

In practice, to enable scalable search, vector databases construct index structures that organize vectors via graphs, trees, hashmaps, or partitions~\cite{HNSW, malkov2018efficient, IVFFlat, bentley1975multidimensional, jayaram2019diskann, LSH}. 
These indices enable Approximate Nearest Neighbor Search (ANNS), which searches only a subset of stored vectors.
ANNS is often controlled by a \textit{global search parameter} that governs \emph{search effort}, i.e., the number of stored vectors scanned per query.
Thus, ANNS improves search performance while potentially missing some true nearest neighbors, compromising search quality~\cite{har2012approximate, ren2020hm, bentley1975multidimensional, hajebi2011fast}. 
Search quality loss is measured by \emph{recall}, defined as the fraction of true nearest neighbors retrieved; recall closer to 1 indicates better search quality~\cite{aumuller2020ann}.


%
\begin{figure}[t!]
  \centering
  \includegraphics[width=0.375\textwidth]
  {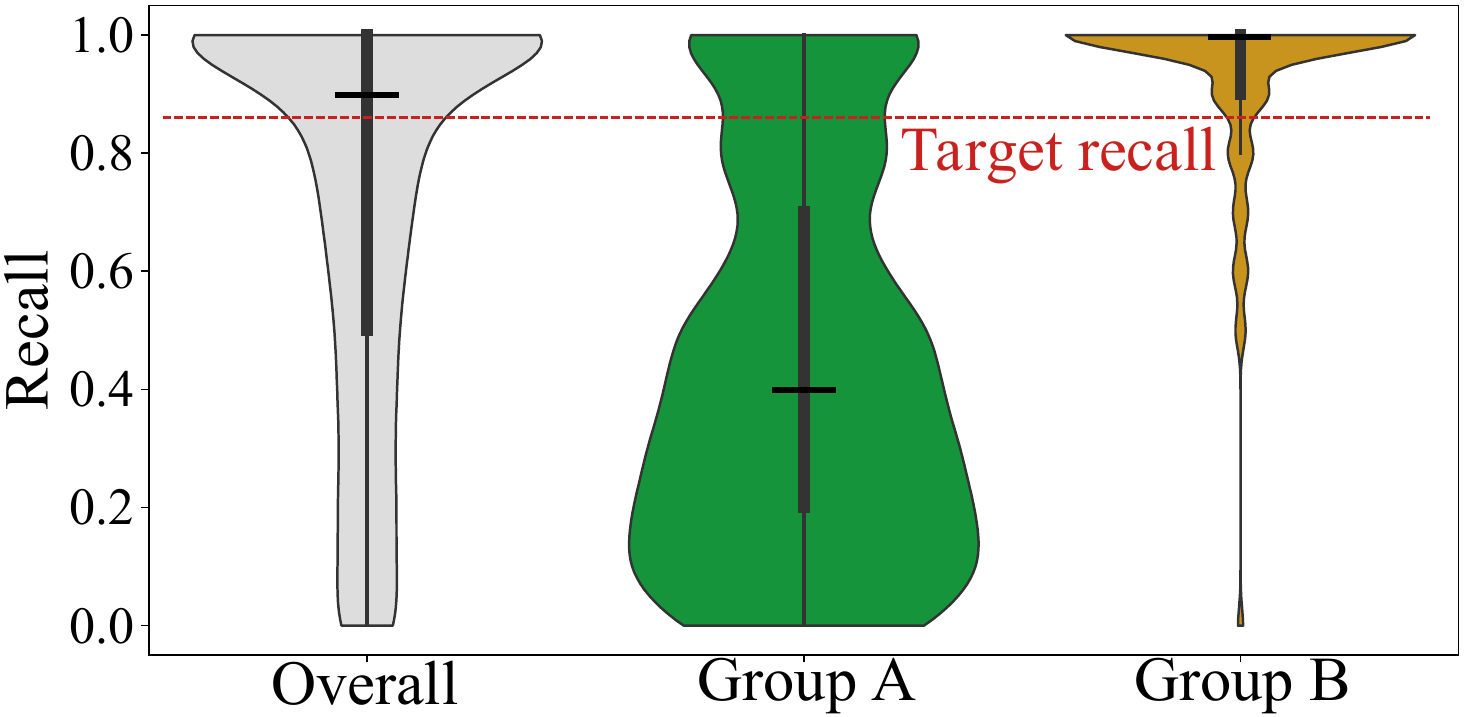}
  \caption{
  Recall distribution across all queries and within two query {\cluster}s corresponding to different regions of the embedding space. 
  Although the overall mean recall meets the target, recall varies substantially across {\cluster}s, indicating uneven search quality across queries.
  }
  \label{fig:fig_recall_violinplot_bob_alice}
\end{figure}

Typically, database operators tune indices to achieve a target mean recall (e.g., 0.85) while maximizing throughput across queries~\cite{ren2020hm, weaviate, wang2025towards}.
We find that optimizing for mean recall can mask a critical problem: search quality can differ substantially across queries even when target recall is met.   

To illustrate recall differences across queries, Fig.~\ref{fig:fig_recall_violinplot_bob_alice} shows overall recall distribution for a \emph{pgvector} database~\cite{pgvector} with a graph-based index~\cite{HNSW} built on a Twitter dataset~\cite{glovedatasets} with \glove-100 embeddings~\cite{pennington2014glove}.
We split the dataset into index and query sets.
While overall mean recall of 0.87 exceeds the 0.85 target, the distribution is wide rather than concentrated, indicating significant difference in recall across queries.

To better study recall differences across queries, we organize stored vectors into \emph{{\cluster}s} based on their location in the embedding space using popular k-means clustering~\cite{KMeans-scikit-learn}.
We show that computing such {\cluster}s offline, independent of the index structure, provides a logical characterization of the embedding space without modifying the index, which enables efficiently studying recall differences across queries.    

Fig.~\ref{fig:fig_recall_violinplot_bob_alice} shows significant recall differences among real-world Twitter queries~\cite{glovedatasets} that naturally map to different {\cluster}s based on their embeddings: queries in {\cluster}~A average 0.21 recall while those in B average 0.99, despite similar cluster sizes.
Thus, even when net mean recall meets the target, search quality can differ greatly across query {\cluster}s.
We show that recall differences persist across index configurations, index types, and embedding models~\cite{schwartz2025important, li2024does}.

Such recall differences have adverse consequences.
Queries that belong to low-recall {\cluster}s experience poor search quality, which hurts (1) user experience, increasing user abandonment and reducing engagement~\cite{al2010review, abualsaud2020effect, wu2018turning, karmaker2017application} and (2) user trust and reliable service operation, hurting revenue~\cite{wang2025towards, chronis2025filtered, diaz2020evaluating, ekstrand2022fairness}. 
Thus, it is crucial to detect and reduce recall differences across queries.

Reducing recall differences is challenging. 
One option is to modify the index offline (e.g., by increasing graph connectivity for low-recall {\cluster}s). 
But, such approaches are expensive and disruptive, making them challenging to deploy in production~\cite{malkov2018efficient}.   
An alternative is to configure the global search parameter that controls \emph{search effort}~\cite{pgvector, HNSW, IVFFlat}.
In current practice, this parameter is set coarsely and the same search effort is applied across all queries.
Thus, increasing this parameter improves recall for queries in low-recall {\cluster}s (e.g., {\cluster}~A), but, also increases search effort for queries that already meet the target recall, reducing net throughput.

Our key insight is that we can reduce recall differences while achieving high throughput by tuning search effort at a finer granularity, i.e., increasing effort for {\cluster}s below target recall and decreasing it for those above it.
Such tuning ``takes from the rich and gives to the poor'' by reallocating computation from high-recall {\cluster}s to low-recall {\cluster}s, reducing recall differences while ensuring high throughput.


Using this insight, we design \rcheck, a lightweight \runtime system that (1) identifies low-recall queries and (2) reduces recall differences while achieving high throughput.
\rcheck introduces several novel design principles.
First, as computing recall differences across individual queries is expensive at modern scales, \rcheck efficiently captures such differences by introducing \emph{per-{\cluster}} recall measurement.
Second, unlike the modern coarse-grained practice of applying uniform search effort across queries, \rcheck dynamically adapts effort \emph{per-{\cluster}} in a way that reduces recall difference while ensuring high throughput. 
Third, for quick adaptation, \rcheck includes control-loop feedback to monitor recall and performance metrics and suitably adapt search effort at run time.
Fourth, \rcheck tunes available search effort parameters, requiring no expensive, intrusive index changes, making it index independent and readily deployable.

We evaluate \rcheck using pgvector~\cite{pgvector}, a widely-used PostgreSQL extension with production-style ANNS~\cite{amazon-pgvector, aumuller2020ann, salesforce-pgvector, google-pgvector}.
Across multiple datasets and index types, 
\rcheck consistently improves recall-throughput trade-off.
At the same target recall, \rcheck improves throughput by \minThroughputAtTargetRecall--\maxThroughputAtTargetRecall\% compared to the best globally-tuned configuration.
At the same throughput, \rcheck tightens the recall distribution, improving mean recall by \minRecallatIsoThroughput--\maxRecallatIsoThroughput\% and enabling \minQueriesMeetingTarget--\maxQueriesMeetingTarget\% more queries to meet target recall target.

In summary, we contribute:
\squishlist
\item A demonstration that ANN recall can differ significantly across queries, despite meeting the target recall.
\item A characterization of this recall difference across index types, index configurations, and embedding models.
\item \rcheck: A light-weight \runtime system that reduces recall differences across queries without hurting throughput.
\item A detailed evaluation of \rcheck, demonstrating that it shifts the recall-performance tradeoff closer to the ideal.

\squishend

\vspace{-4mm}
\section{Background}
\label{sec:background}
Vector databases, such as pgvector~\cite{pgvector}, FAISS~\cite{douze2024faiss}, Milvus~\cite{wang2021milvus}, and Weaviate~\cite{weaviate} store embedding vectors at scale~\cite{ren2020hm}. 
For a query vecto
r, the database retrieves most-similar stored vectors under a distance metric (e.g., Euclidean distance)~\cite{har2012approximate, ren2020hm}. 

Exact similarity search compares the query vector against all stored vectors, which is computationally infeasible at modern scales~\cite{har2012approximate, ren2020hm, clarkson1994algorithm}.
To improve scalability, these databases use Approximate Nearest Neighbor Search (ANNS)~\cite{har2012approximate}, which trades search accuracy for lower latency and higher throughput. 
ANNS stores vectors using index structures such as graphs, trees, hashmaps, or partitions, enabling searching only a subset of stored vectors for a query~\cite{HNSW, malkov2018efficient, IVFFlat, bentley1975multidimensional, jayaram2019diskann, LSH}. 
Thus, ANNS may miss some true nearest neighbors, hurting search quality, which is quantified using \emph{recall}, the fraction of true neighbors retrieved~\cite{aumuller2020ann}. 
In practice, operators configure index parameters to achieve a target \emph{mean recall} while maximizing throughput~\cite{ren2020hm, weaviate, wang2025towards, aumuller2020ann}.

Widely-used ANNS indexes include Hierarchical Navigable Small Worlds ({\hnsw}) and {\ivfflat}. 
{\hnsw} organizes vectors as a multi-layer graph which it greedily traverses to find nearest neighbors~\cite{HNSW, malkov2018efficient}. 
{\ivfflat} partitions vectors into clusters and searches clusters closest to the query~\cite{IVFFlat}.
These ANNS techniques expose tunable parameters that govern index construction and \runtime search effort, which together determine search quality vs.\ throughput tradeoff~\cite{HNSW, IVFFlat}.

In {\hnsw}, build-time parameters determine the graph structure: \texttt{M} controls the number of connections per stored vector and \texttt{ef\_construction} controls how many candidates are considered when selecting these connections during insertion. 
At \runtime, \texttt{ef\_search} determines how many stored vectors are scanned per query. 
In {\ivfflat}, \texttt{nlist} dictates the number of partitions created during index construction.
\texttt{nprobes} controls how many partitions are searched at \runtime~\cite{HNSW, IVFFlat}. 
Increasing these parameters typically improves recall by exploring the embedding space more at the cost of higher latency and lower throughput (in Queries Per Second---QPS); decreasing them has the opposite effect~\cite{aumuller2020ann}. 

Build-time parameters (\texttt{M}, \texttt{ef\_construction}, \texttt{nlist}) require index reconstruction to modify. 
\Runtime parameters which determine \textit{search effort} (\texttt{ef\_search} and \texttt{nprobes}) can be adjusted dynamically without index reconstruction. 
In practice, such \runtime parameters are typically configured globally and applied uniformly across all queries~\cite{aumuller2020ann, wang2021milvus, douze2024faiss}. 

\section{Studying Recall Differences Across Queries to a Vector Database}
\label{sec:motivation}

We demonstrate recall differences across queries to a vector database, motivating why it is key to reduce such differences. 
We show that such differences persist even when we change the embedding model, index type, index construction, and global search effort, motivating our solution, \rcheck.

\textbf{Demonstrating recall differences across queries.}
Modern vector databases are tuned to achieve a target mean recall while maximizing throughput~\cite{ren2020hm, weaviate, wang2025towards,aumuller2020ann}. 
We find that when optimizing for the mean recall alone, recall can differ significantly across embedding space regions even when the target mean recall is met.   


To characterize recall differences across queries, we organize stored vectors into \emph{{\cluster}s} based on their location in the embedding space using popular k-means clustering~\cite{KMeans-scikit-learn} during offline index construction. 
The \textit{k} centroids represent distinct embedding space regions.
We map each query to the {\cluster} whose centroid is nearest to it using the same distance metric used by the underlying index.
Such {\cluster}ing helps us conveniently analyze whether queries that map to a certain embedding space region face significantly different recall compared to queries that map to another region.


Fig.~\ref{fig:fig_recall_violinplot_bob_alice} shows the recall distribution for two such {\cluster}s ({\cluster}~A and {\cluster}~B).
Experimental setup is the same as the one described in \S\ref{sec:intro}.
Despite using identical search configurations and covering comparable portions of the dataset, {\cluster}~A achieves a mean recall of only 0.21, while {\cluster}~B achieves 0.99.
This example shows that queries targeting different regions of the embedding space can experience significantly different recall.

\textbf{Why identify and reduce recall differences?}
The recall differences illustrated above have consequences for both users and the vector search system.
Queries in low-recall regions miss relevant results, even when the overall mean recall appears to satisfy the target recall. 
As noted in in \S~\ref{sec:intro}, recall is correlated with user satisfaction and can lead to user disengagement and/or negatively impact the end application that is using the vector search system.

An interesting observation is that these recall differences can align with semantic groups.
For example, in Fig.\ref{fig:fig_recall_violinplot_bob_alice}, post-hoc analysis shows that over 90\% of the words in the low-recall {\cluster} A correspond to Arabic tokens, while over 92\% of the words in a high-recall {\cluster} B are Latin-script tokens. 
\jmp{Important to note if both groups have similar amount of data, which could be quantified as the the number of tweets or words.}
We emphasize that this observation does not imply inherent superiority of one language over another. 
Rather, it highlights that ANN index behavior can interact with embedding geometry in ways that disproportionately affect specific semantic groups. 
When such {\cluster}-level recall differences align with meaningful query groups, low recall is no longer an isolated issue affecting individual queries, but a systematic problem affecting entire classes of users. 

\textbf{Why do potential offline solutions fall short?}
A possible solution is to reduce recall differences across queries through offline solutions, such as, switching embedding models, using another index or rebuilding the index with more aggressive construction parameters.
However, these solutions are insufficient.
They are expensive and disruptive in production: they require recomputing embeddings and/or rebuilding the index, increasing computational cost and memory footprint, and complicating deployment in production environments~\cite{malkov2018efficient}.
Moreover, we find that recall differences can persist even after applying these solutions, and when they do help, it often comes at the cost of significant throughput degradation.
We show this through the case studies below. 

\begin{figure}[t]
\centering 
\subfloat[][\glove-200 with \hnsw]{
  \includegraphics[clip,width=0.23\textwidth]{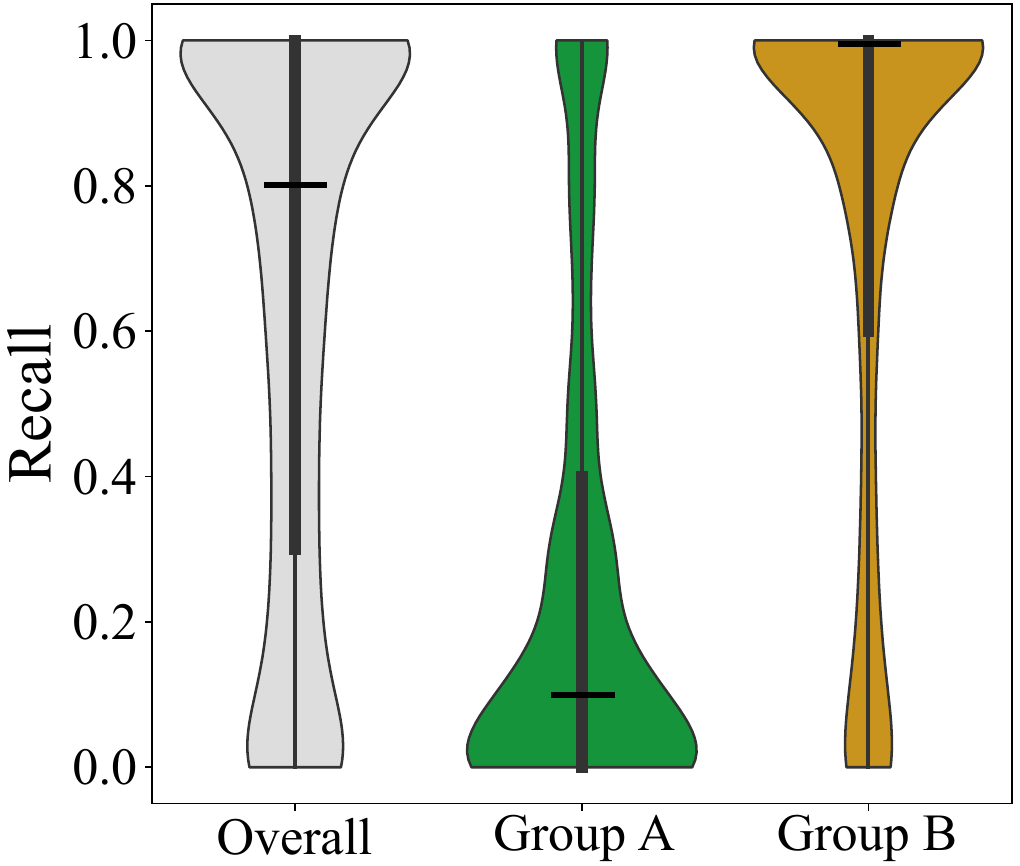}
\label{fig:fig_glove_200_embedding_distribution_breakdown}
}
\subfloat[][\glove-100 with \ivfflat]{
\includegraphics[clip,width=0.23\textwidth]
{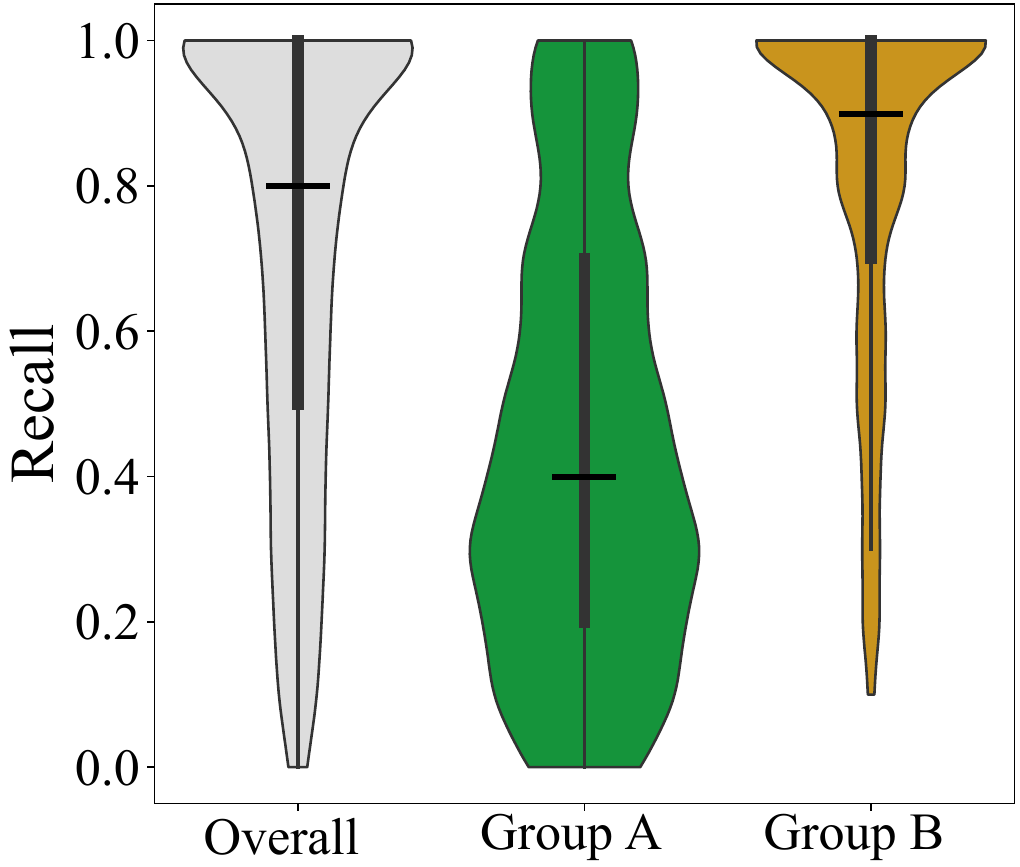}
\label{fig:fig_glove_100_ivfflat_distribution_breakdown}
}
\caption{
Recall distributions across all queries and across the low-recall and high-recall {\cluster}s for two ANN index configurations: (a) \glove-200 with {\hnsw} and (b) \glove-100 with {\ivfflat}.
Even when the overall average recall differs across configurations, recall remains widely spread, with substantial gaps between low-recall and high-recall {\cluster}s.
}
\vspace{-3mm}
\end{figure}

\emph{Switching embedding model.}
One might try to address recall differences by switching to a different embedding model.
To test this, we replace the \glove-100 embedding model with \glove-200~\cite{pennington2014glove}, a different model that represents each word as a 200-dimensional vector instead of 100 dimensions, on the same Twitter dataset~\cite{glovedatasets} used in Fig.\ref{fig:fig_recall_violinplot_bob_alice}. 

Fig.~\ref{fig:fig_glove_200_embedding_distribution_breakdown} shows recall distribution for this higher-dimensional embedding model.
Similar to \glove-100, the overall recall distribution is wide, with significant variation across queries.
When we partition the embedding space into {\cluster}s and examine recall at the {\cluster} granularity, we see that recall differences persist across {\cluster}s.
For example, two {\cluster}s of similar size (3\% of the dataset each) exhibit vastly different recall: {\cluster} B achieves mean 0.81 (25th-75th percentile: 0.6-1.0) vs.\ {\cluster} A's mean of 0.26 (25th-75th percentile: 0-0.4).
We do not claim that all embedding models exhibit identical recall behavior. 
However, this case study shows that recall differences across {\cluster}s are not artifacts of a specific embedding model and they can persist across commonly used embedding representations, even if their magnitude varies.

\emph{Switching ANN index types.}
One might hypothesize that recall differences are specific to a particular ANN index type.
To test this, we replace the graph-based {\hnsw} index~\cite{HNSW} with {\ivfflat}~\cite{IVFFlat}, a partition-based index, on the same Twitter dataset with \glove-100 embeddings as in Fig.~\ref{fig:fig_recall_violinplot_bob_alice}.
Fig.~\ref{fig:fig_glove_100_ivfflat_distribution_breakdown} shows the resulting recall distribution.
We observe that, although IVFFlat changes the absolute recall values, the overall distribution remains wide.
When we examine recall at the {\cluster} granularity, these differences persist: {\cluster}s of comparable size exhibit substantially different recall behavior, e.g., {\cluster}~A: mean 0.44, 25th-75th percentile: 0.2-0.7; {\cluster}~B: mean 0.82, 25th-75th percentile: 0.7-1.0.
This demonstrates that recall differences are not specific to an ANN index  and can persist across different index types.

\begin{figure}[t]
\centering 
\subfloat[][{\hnsw}: Varying \texttt{M}]{
  \includegraphics[clip,width=0.23\textwidth]{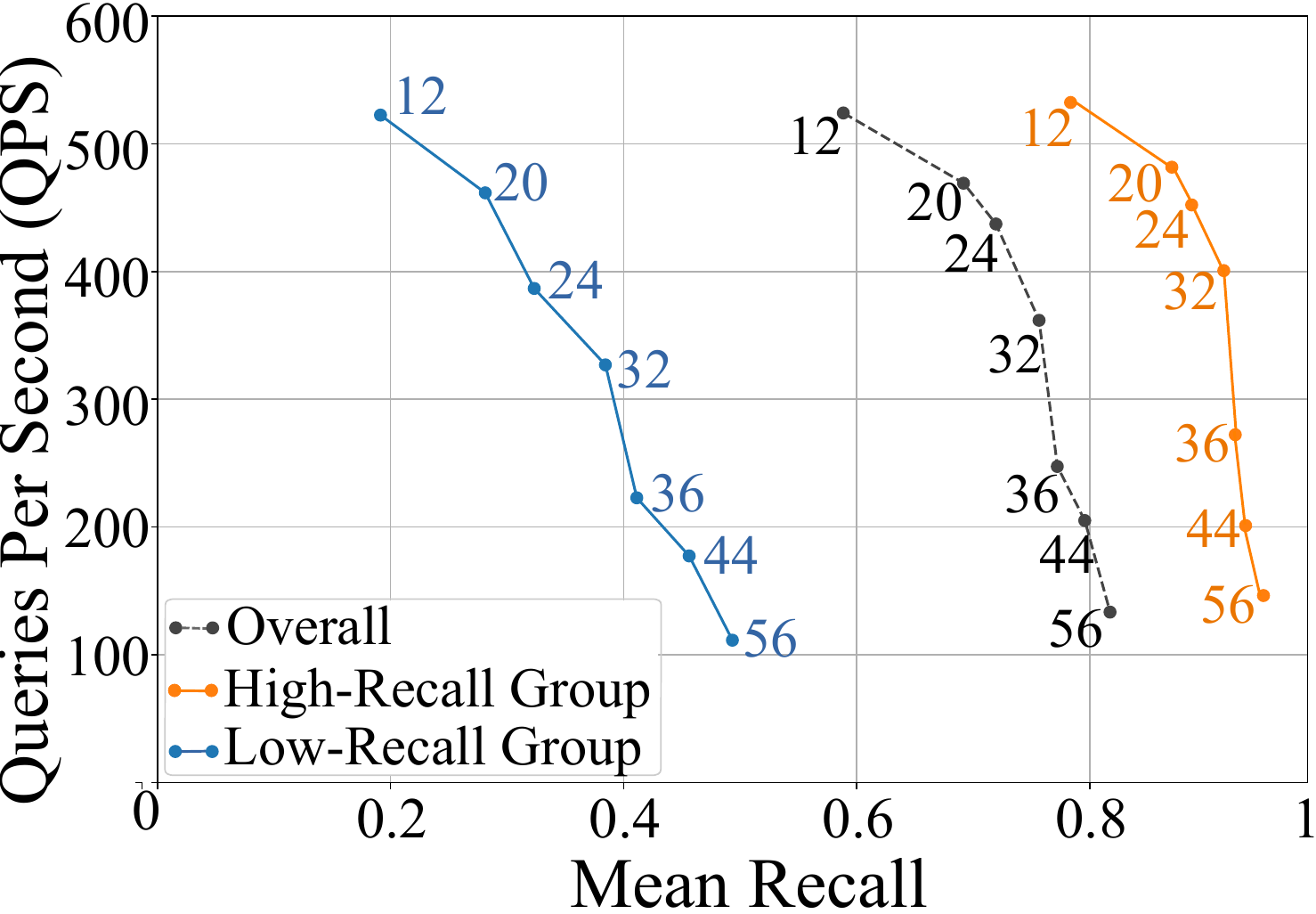}
  \label{fig:fig_glove100_pareto_frontier_all_M_clusters_avg_recall_vs_qps_raw}
}
\subfloat[][{\ivfflat}: Varying \texttt{nlist}]{

\includegraphics[clip,width=0.231\textwidth]{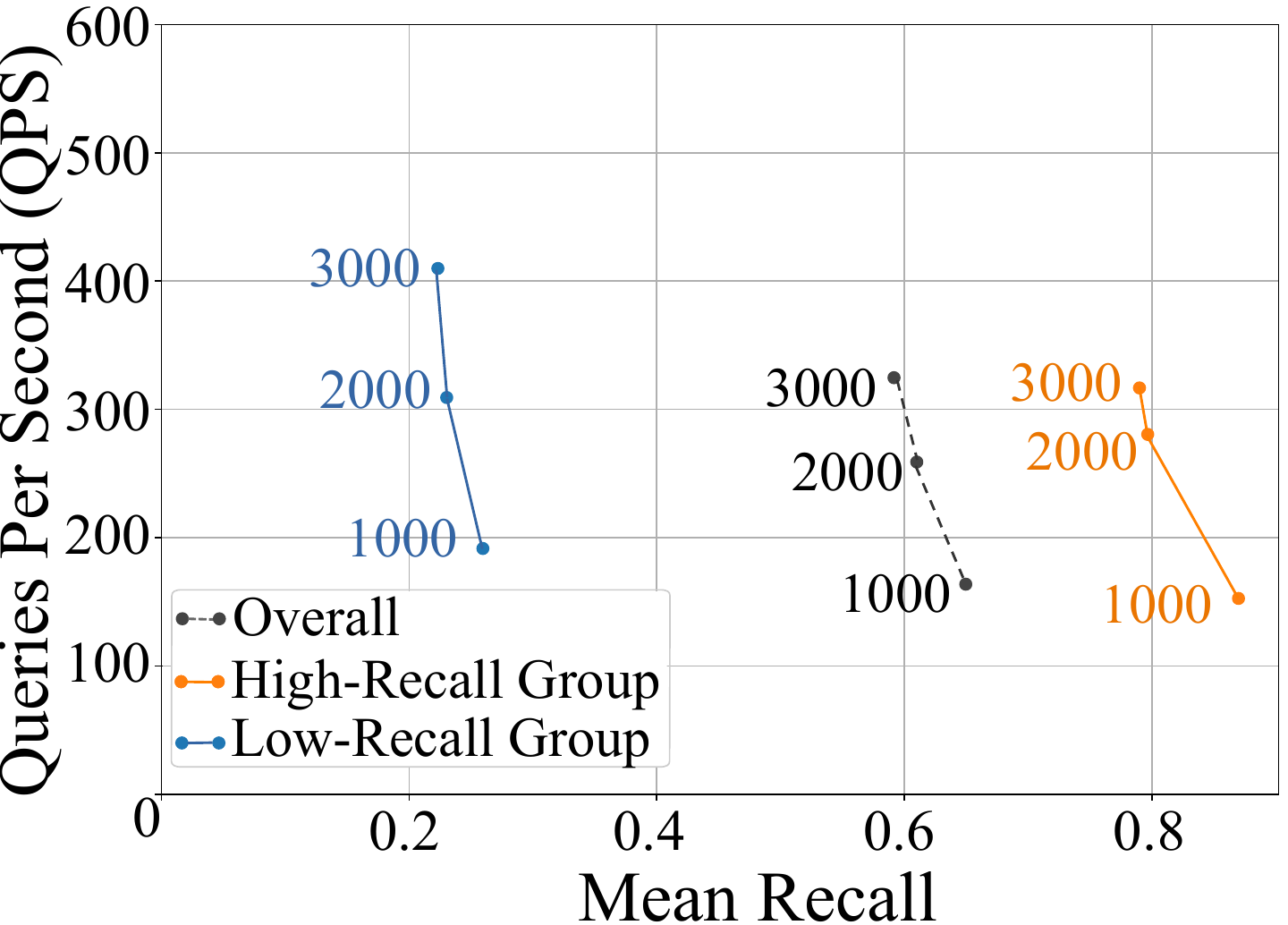}
\label{fig:fig_glove100_pareto_frontier_all_nlist_clusters_avg_recall_vs_qps_raw}
}
\caption{
QPS–recall trade-offs for the low-recall and high-recall {\cluster}s and the overall distribution as index construction parameters are varied: (a) {\hnsw} with fixed \texttt{ef\_construction}=200 and \texttt{ef\_search}=20 while varying \texttt{M}, and (b) {\ivfflat} with fixed \texttt{probes}=4 while varying \texttt{nlist}.
Although tuning these parameters improves recall for the low-recall {\cluster}, it 
degrades 
QPS; also, the same parameter changes affect recall improvement differently across {\cluster}s.
}
\vspace{-3mm}
\end{figure}

\emph{Tuning index construction parameters.}
We examine whether tuning index construction parameters can mitigate recall differences.
To test this, we use the 
Twitter dataset~\cite{glovedatasets}, \glove-100 embedding model, and {\cluster}ing as in Fig.~\ref{fig:fig_recall_violinplot_bob_alice}.

For {\hnsw}, we vary \texttt{M} (the number of connections per vector), which controls graph density.
Higher \texttt{M} improves recall at the cost of increased memory and reduced throughput.
Fig.~\ref{fig:fig_glove100_pareto_frontier_all_M_clusters_avg_recall_vs_qps_raw} plots QPS vs. mean recall for two example {\cluster}s, one low-recall and one high-recall, alongside the overall distribution.
Although higher \texttt{M} improves recall both overall and within individual {\cluster}s, significant recall differences persist.
Moreover, {\cluster}s do not benefit equally from increasing \texttt{M}: at \texttt{M}=12, the low-recall {\cluster} achieves 0.19 vs. 0.78 for the high-recall {\cluster}; at \texttt{M}=56, these differences persist at 0.50 and 0.95, respectively.
Increasing \texttt{ef\_construction} (the search effort during index building) follows a similar trend: absolute recall improves, but recall differences remain.

We observe the same pattern for {\ivfflat} (Fig.~\ref{fig:fig_glove100_pareto_frontier_all_nlist_clusters_avg_recall_vs_qps_raw}).
We vary \texttt{nlist} (the number of partitions) while keeping \texttt{nprobes} (partitions searched per query) fixed
.
Increasing \texttt{nlist} creates smaller partitions; however, with constant \texttt{nprobes}, overall recall decreases as nearest neighbors increasingly fall outside the searched partitions.
Recall differences across {\cluster}s persist nonetheless: at \texttt{nlist}=1000, 
low-recall {\cluster} achieves mean recall of 0.27 vs. 0.95 for the high-recall {\cluster}; at \texttt{nlist}=3000, mean recall is 0.22 vs. 0.89, respectively.

This illustrates that tuning index construction parameters does not eliminate recall differences and even when such tuning reduces recall differences, it often comes at the cost of significantly reduced throughput.




\begin{figure}[t]
\centering 
\subfloat[][HNSW: Varying \texttt{ef\_search}]{
  \includegraphics[clip,width=0.23\textwidth]{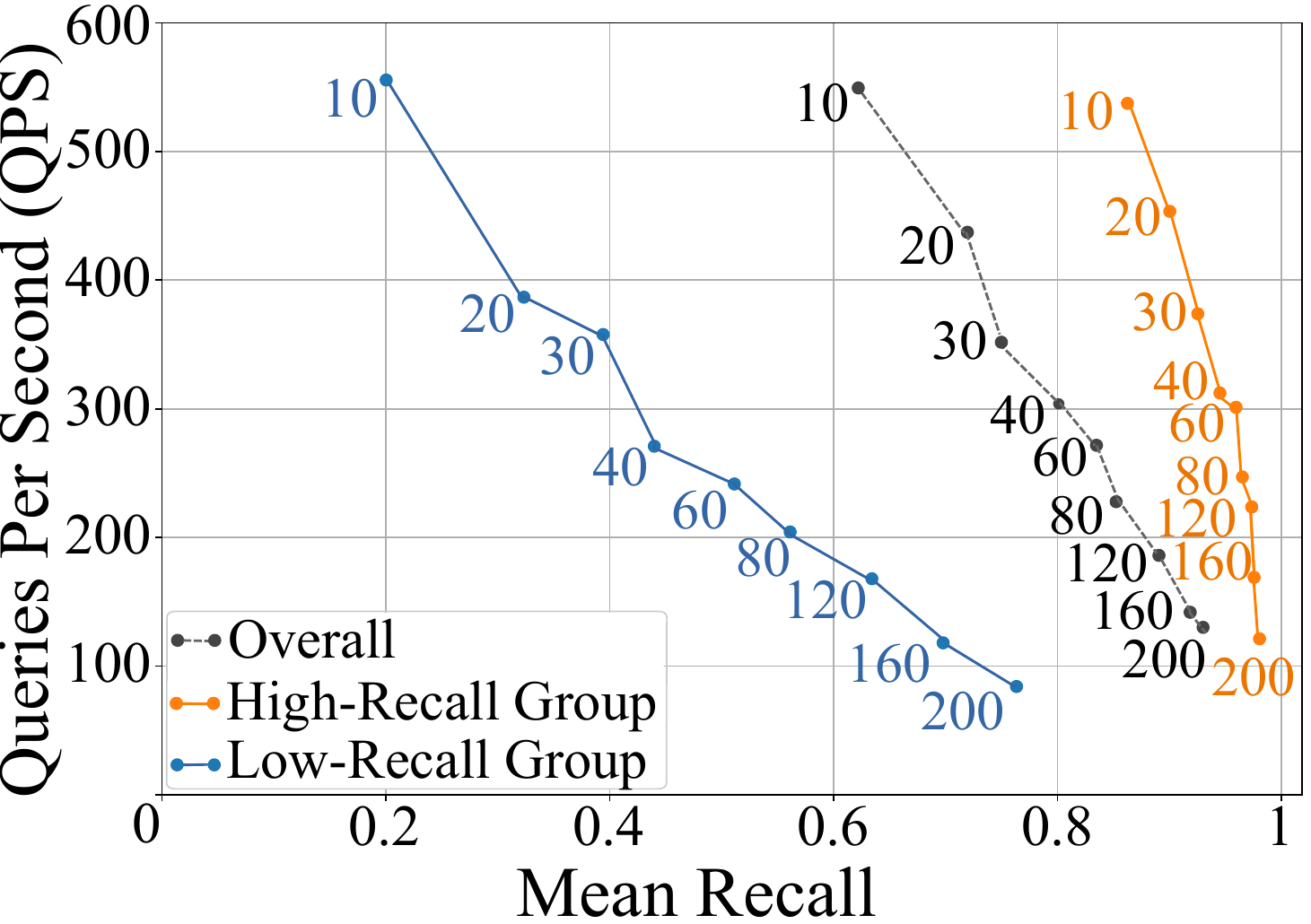}
  \label{fig:fig_glove100_pareto_frontier_all_clusters_avg_recall_vs_qps_raw}
}
\subfloat[][IVFFlat: Varying \texttt{probes}]{

\includegraphics[clip,width=0.23\textwidth]{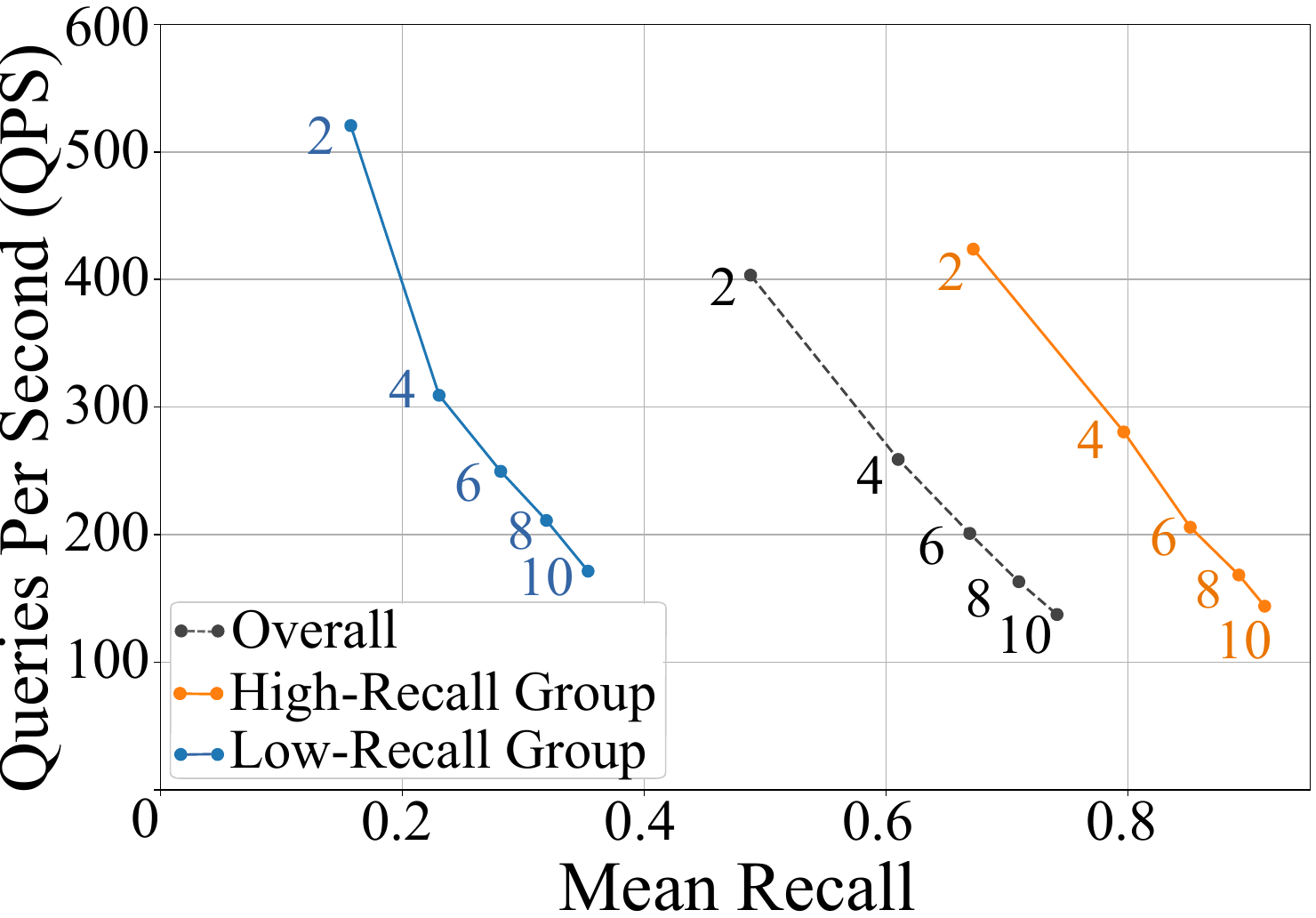}
\label{fig:fig_glove100_pareto_frontier_all_probes_clusters_avg_recall_vs_qps_raw}
}
\caption{
QPS–recall trade-offs for the lowest- and highest-performing {\cluster}s and the overall distribution as {\hnsw}’s \texttt{ef\_search} (a) and {\ivfflat}’s \texttt{probes} (b) are varied.
Increasing 
search effort improves recall for low-recall {\cluster}s but also incurs unnecessary work for high-recall {\cluster}s, leading to degrading throughput; 
also, identical increases in search effort yield different recall improvements across {\cluster}s.}
\vspace{-3mm}
\end{figure}

\textbf{Why do typical \runtime solutions fall short?}
At \runtime, operators can tune a \emph{search effort} parameter (e.g., \texttt{ef\_search} for HNSW or \texttt{nprobes} for IVFFlat), but this is typically configured for the entire index.
Increasing search effort helps low-recall {\cluster}s, but it also increases work for {\cluster}s that already achieve satisfactory recall, leading to unnecessary computation and reduced net throughput.

Fig.~\ref{fig:fig_glove100_pareto_frontier_all_clusters_avg_recall_vs_qps_raw} illustrates this trade-off for HNSW: as \texttt{ef\_search} increases, recall for low-recall {\cluster}s improves, but net throughput degrades due to extra work on high-recall {\cluster}s.
Moreover, increasing search effort can benefit each {\cluster} differently.
At \texttt{ef\_search}=20, the low-recall {\cluster} achieves recall of 0.32 vs. 0.90 for the high-recall {\cluster}; at \texttt{ef\_search}=160, these become 0.70 and 0.97, respectively.

Fig.~\ref{fig:fig_glove100_pareto_frontier_all_probes_clusters_avg_recall_vs_qps_raw} shows a similar pattern for IVFFlat when we increase \texttt{nprobes}.
At \texttt{nprobes}=2, the low-recall {\cluster} achieves recall of 0.18 vs. 0.67 for the high-recall {\cluster}; at \texttt{nprobes}=10, these become 0.38 and 0.97, respectively.

This underscores that global \runtime tuning cannot eliminate recall differences and improvements for low-recall {\cluster}s result in unnecessary computation for {\cluster}s that already perform well, reducing overall throughput.


\textbf{Takeaway.}
Both ANN index construction and {\runtime} tuning fall short in reducing recall differences across queries without significantly hurting overall throughput.
This motivates a finer-grained approach: selectively adjusting search effort per {\cluster} based on recall profiles by concentrating more effort for low-recall {\cluster}s and avoiding unnecessary work for {\cluster}s that already meet the target recall.
We design \rcheck to enable this finer-grained approach. 
\section{RCheck}
\label{sec:design}
Building on the observations in §\ref{sec:motivation}, we introduce \rcheck, a light-weight \runtime adaptation system that detects {\cluster}s whose recall diverges from the target recall and applies fine-grained, \cluster-specific search adaptations to reduce these differences.
Before detailing \rcheck’s modules, we 
describe design goals that shape \rcheck's architecture and workflow.


\subsection{\emph{\rcheck} Design Goals}
\vspace{-0.5em}
\rcheck's goal is to balance recall across queried regions of the embedding space at {\runtime} by increasing search effort for low-recall {\cluster}s while reducing unnecessary work for {\cluster}s whose recall already exceed the target.
To achieve this goal, we design \rcheck around three principles:

\textbf{{\Runtime}, non-intrusive operation.}
\rcheck operates entirely at \runtime without modifying embeddings, rebuilding the ANN index, or changing the database architecture. 
This ensures \rcheck is easy to deploy in production and avoids expensive offline reconfiguration.

\textbf{Fine-grained search effort adaptation.}
Common index configuration practices to address recall differences across queries fall short. 
They either cause queries that map to some regions of the embedding space to face low recall or lead to unnecessary work for queries that map regions that achieve high recall even with low search effort, degrading overall throughput.
\rcheck avoids this by adjusting search effort per {\cluster} at \runtime, guided by each {\cluster}'s recall profile.

\textbf{Balancing the recall vs. throughput tradeoff.}
Improving recall can hurt throughput.
To balance this, \rcheck lets operators define performance constraints (e.g., throughput), 
then redistributes search effort across {\cluster}s: reducing effort for {\cluster}s exceeding the target recall and increasing it for those falling short, balancing recall while staying within the operator-defined performance budget.

\begin{figure}[t!]
  \centering
  \includegraphics[width=0.47\textwidth]{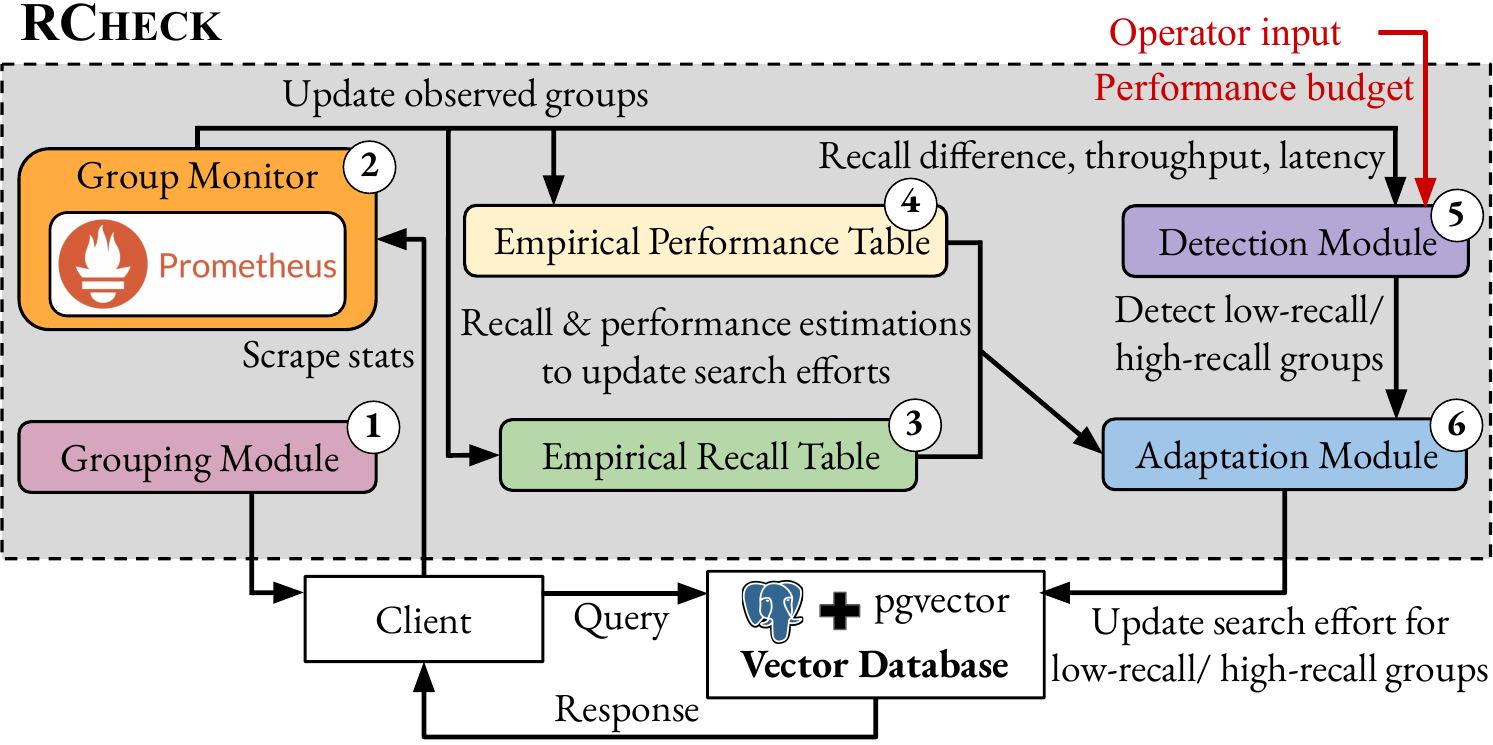}
  \caption{\rcheck system design overview.
  }
  \label{fig:fig_rcheck_design_overview}
\end{figure}

\subsection{\emph{\rcheck} High-level Overview}
\label{sec:desgn-highlevel}
\rcheck operates as a feedback-driven control-loop, as shown in Fig.~\ref{fig:fig_rcheck_design_overview}.
At {\runtime}, it continuously monitors recall and performance metrics, including throughput and tail latency, for each {\cluster}, maintaining light-weight empirical tables that summarize how recall and performance respond to changes in search effort.
Periodically, \rcheck uses these tables to adjust per-{\cluster} search effort, correcting recall deviations from the target while staying within the throughput and tail latency budget defined by the operator.



To enable fine-grained control, \rcheck organizes queries into \emph{{\cluster}s}, which serve as the units of monitoring and adaptation.
Per-query adaptation would be ideal but is computationally expensive: tracking individual query behavior and computing ground truth for every query is impractical at scale.
{\Cluster}s provide a practical middle ground by grouping queries that map to the same region of the embedding space (\S\ref{sec:motivation}).
{\Cluster}s are defined at index construction time using k-means clustering~\cite{lloyd1982least} and are purely logical and impose no changes to the underlying index structure.


We now describe \rcheck's run-time adaptation's stages.

\textbf{{\Cluster} query assignment.}
At \runtime, \rcheck assigns each incoming query to a {\cluster} via nearest-centroid lookup.

\textbf{Per-\cluster\ monitoring.}
\rcheck operates in fixed-length monitoring intervals.
During each monitoring interval, \rcheck samples queries from every {\cluster} and performs exact searches in parallel to obtain true neighbors (i.e., ground truth).
Comparing these with the ANNS results estimates per-{\cluster} recall.
\rcheck also records per-{\cluster} throughput and tail latency observed during the interval.

\textbf{Per-{\cluster} adaptation.}
At the end of each monitoring interval, \rcheck adjusts search effort as follows:

\emph{(i) Empirical table update.}
Using per-{\cluster} statistics collected during the monitoring interval, \rcheck updates two light-weight summaries for each {\cluster}:
a \textit{recall table} recording how recall varies with search effort, and
a \textit{performance table} recording the corresponding throughput and latency.
These tables are updated incrementally at runtime.

\emph{(ii) Recall difference detection.}
For each {\cluster}, \rcheck computes the difference between the recall observed in the current monitoring interval and the target recall.
{\Cluster}s are then ranked by the recall difference magnitude, so those furthest from the target are considered first for adaptation.

\emph{(iii) Search effort adaptation.}
\rcheck processes {\cluster}s in the ranked order, determining whether to increase, decrease, or maintain each {\cluster}'s search effort.
For {\cluster}s below target recall, \rcheck consults the recall table to identify the smallest search effort increase estimated to shift recall towards the target.
For {\cluster}s well above target, \rcheck considers reducing search effort only if the recall table indicates recall will remain above target and the performance table estimates meaningful throughput savings.

\rcheck evaluates each proposed change against the operator-defined performance budget, accounting for all previously accepted adaptations.
Adaptations that fit within the budget are accepted; others are skipped.
By processing {\cluster}s in ranked order, \rcheck ensures that those with the largest recall deficits receive more search effort first, while those with the largest recall surplus contribute throughput savings.

\emph{(iv) Applying the search effort adaptation.}
\rcheck applies the accepted search effort adaptation to the ANN index, which takes effect immediately.
The next monitoring interval then begins, during which, \rcheck observes how the updated search efforts affect recall and throughput.
This feedback informs the next round of search effort adaptations.
\subsection{\emph{\rcheck} Modules}
\label{sec:design:modules}
We now describe the modules that implement \rcheck's workflow (\S\ref{sec:desgn-highlevel}), as illustrated in Fig.~\ref{fig:fig_rcheck_design_overview}.

~{\large \textcircled{\normalsize {1}}}~\textbf{\GroupModule}:
It creates {\cluster}s at offline index construction time and maps incoming queries to them at runtime.
By default, \rcheck partitions the embedding space using k-means clustering~\cite{lloyd1982least} over indexed vectors, producing $k$ centroids that represent distinct regions of the 
embedding 
space.
Importantly, the grouping is purely logical and does not modify the underlying ANN index structure or internal database data structures.

At \runtime, each incoming query is mapped to a {\cluster} using nearest-centroid lookup: the query embedding is compared against the $k$ centroids, and the query is mapped to the {\cluster} with the smallest distance, enabling fine-grained, per-{\cluster} monitoring and search effort adaptation.



~{\large \textcircled{\normalsize {2}}}~\textbf{\GroupMonitor}:
It continuously collects per-{\cluster} statistics during each monitoring interval.
For each {\cluster}, it samples a small number of queries and executes exact searches in parallel to obtain the true nearest neighbors (i.e., ground-truth).
It then calculates the recall by comparing these ground-truth results with the ANNS results returned along the normal query serving path.
The \groupMonitor also records per-{\cluster} density (i.e., query count), throughput, and p99 tail latency observed during the monitoring interval.
All monitoring and exact-search executions run asynchronously and do not block query processing.

~{\large \textcircled{\normalsize {3}}},{\large \textcircled{\normalsize {4}}}~\textbf{Empirical recall and performance tables.}
For each {\cluster}, \rcheck maintains two empirical tables:
(i) a \textit{recall table} that records observed recall values under different search effort settings and
(ii) a \textit{performance table} that records the corresponding throughput 
and p99 latency.
Each table stores <search effort, metric> pairs collected during monitoring intervals, capturing how changes in search effort affect recall and performance at {\cluster} granularity.

\rcheck uses these tables to estimate the direction and magnitude of change in recall or throughput as search effort is adjusted, enabling incremental, gradient-like updates at {\runtime} without fitting an explicit model.
The tables become more accurate as observations accumulate.
When historical per-{\cluster} statistics are available from a previous execution, \rcheck can optionally load them to initialize the tables and reduce cold-start overhead.
\begin{algorithm}[htbp]
\caption{\rcheck Intervention}
\label{alg:rcheck_mitigation}

\KwIn{ranked {\cluster}s $\mathcal{G}$; recall table $T_r$; performance table $T_p$;
target recall $r^*$; throughput budget $B_t$; latency threshold $L_{max}$}

$\mathcal{A} \gets \emptyset$\tcp*[r]{Accepted adaptations}
$b^{\text{rem}} \gets B_t$\tcp*[r]{Remaining throughput budget}

\ForEach{{\cluster} $g \in \mathcal{G}$ in order of $|r_g - r^*|$}{
    \If{$|r_g - r^*| < \epsilon$}{
        \textbf{continue}\tcp*[r]{Within tolerance}
    }
    
    $(\Delta e, \Delta b) \gets \textsc{GetEffort}(g, r^*, T_r, T_p, b^{\text{rem}}, L_{max})$\;
    
    \If{$\Delta e \neq 0$}{
        $\mathcal{A} \gets \mathcal{A} \cup \{(g, \Delta e)\}$\;
        $b^{\text{rem}} \gets b^{\text{rem}} - \Delta b$\;
    }
}

\textsc{ApplyAdaptation}$(\mathcal{A})$\;

\vspace{0.5em}
\hrule
\vspace{0.5em}

\SetKwProg{Fn}{Function}{:}{}
\Fn{\textsc{GetEffort}$(g, r^*, T_r, T_p, b^{\text{rem}}, L_{max})$}{
    $\Delta e_r \gets \textsc{RecallDrivenAdaptation}(g, r^*, T_r)$\;
    
    \uIf{$r_g < r^*$}{
        \tcp{Low-recall {\cluster}: Increase effort}
        $\Delta e_b \gets \textsc{BudgetDrivenLimit}(g, T_p, b^{\text{rem}})$\;
        $\Delta e_l \gets \textsc{LatencyDrivenLimit}(g, T_p, L_{max})$\;
        $\Delta e \gets \min(\Delta e_r, \Delta e_b, \Delta e_l)$\;
        
        \If{$\Delta e \le 0$}{
            \Return{$(0, 0)$}\tcp*[r]{Violates budget}
        }
    }
    \Else{
        \tcp{High-recall {\cluster}: Decrease effort}
        $\Delta e \gets \Delta e_r$

        \If{$\Delta e < \lambda$}{
            \Return{$(0, 0)$}\tcp*[r]{Violates target}
        }
        
        \If{$\textsc{ThroughputGain}(g, \Delta e, T_p) < \gamma$}{
            \Return{$(0, 0)$}\tcp*[r]{Gain too small}
        }
    }
    
    $\Delta b \gets \textsc{EstimateCost}(g, \Delta e, T_p)$\;
    \Return{$(\Delta e, \Delta b)$}\;
}
\end{algorithm}

{\large \textcircled{\normalsize {5}}}~\textbf{\DetectionModule}:  
At the end of each monitoring interval, this module compares the measured recall of each \cluster\ with the target recall.  
It computes the recall differences and ranks {\cluster}s by difference magnitude, ensuring those furthest from the target are considered first for adaptation.
{\Cluster}s below the target are marked for increased search effort; those above target are candidates for reduction.

{\large \textcircled{\normalsize {6}}}~\textbf{\MitigationModule}:  
It balances recall across the queried portion of the embedding space by adjusting per-{\cluster} search effort to bring each {\cluster} closer to the target recall, subject to operator-defined performance budgets.
Algorithm~\ref{alg:rcheck_mitigation} summarizes this process.

The module processes {\cluster}s in ranked order, prioritizing those with the largest recall difference from the target $r^*$ (line 3).
{\Cluster}s whose recall falls within tolerance $\epsilon$ of the target require no adaptation and are skipped (lines 4--5).

For each remaining {\cluster}, the \textsc{GetEffort} function determines the appropriate search effort adaptation $\Delta e$ and its associated estimated throughput cost $\Delta b$ (line 6).

The \textsc{GetEffort} function first computes the recall-driven adaptation $\Delta e_r$ (line 12) using the recall table $T_r$.
$T_r$ records, for each {\cluster}, the observed recall at previously used effort levels.
From these observations, the function estimates the recall slope for the {\cluster}: the expected change in recall per unit change in effort.
The recall-driven adaptation is then computed as the effort change needed to close the gap between the current recall $r_g$ and the target $r^*$.
For low-recall {\cluster}s, this adaptation is positive (increase effort); for high-recall {\cluster}s, it is negative (decrease effort).
The subsequent logic determines whether this adaptation is acceptable given the performance constraints.

For low-recall {\cluster}s, the function computes two additional limits (lines 14--15):
the budget-driven limit $\Delta e_b$, representing the maximum effort increase that the remaining throughput budget can accommodate, and
the latency-driven limit $\Delta e_l$, representing the maximum increase that keeps the {\cluster}'s mean latency below the threshold $L_{max}$.
The function selects the minimum of these three candidates, ensuring the adaptation improves recall as much as possible without violating either limit (line 16).
If no positive adaptation is feasible, the {\cluster} is skipped (lines 17--18).

For high-recall {\cluster}s, the function \textsc{GetEffort} considers reducing effort to avoid unnecessary work and reclaim throughput budget.
It uses the recall-driven adaptation directly, which is negative in this case (line 20).
However, not all search effort reductions are worthwhile: if the reduction magnitude is smaller than the threshold $\lambda$, the adaptation is skipped (lines 21--22). 
This prevents reductions that would bring recall too close to the target, preserving a safety margin against estimation error.
Moreover, if the estimated throughput gain falls below the threshold $\gamma$, the reduction is skipped (lines 23--24).
This prevents unnecessary adaptations that yield negligible throughput benefit.

For accepted adaptations, the function \textsc{GetEffort} estimates the throughput cost $\Delta b$ from the performance table $T_p$ (line 25).
The cost is positive for search effort increases (consuming budget) and negative for effort decreases (reclaiming budget).
Accepted adaptations are collected, and the remaining budget is updated accordingly (lines 7--9).
After processing all {\cluster}s, the adaptations are applied atomically (line 10).
Reclaimed budget from high-recall {\cluster}s becomes available for subsequent low-recall {\cluster}s, enabling the \mitigationModule to redistribute search effort toward regions of the embedding space that need it most.
\subsection{Implementation}
\label{sec:impl}
We implement \rcheck in $\sim$2500 lines of Python code on top of the ANN Benchmarking framework~\cite{aumuller2020ann}.
Our implementation requires no modification to the underlying vector database or ANN index.
\rcheck has four components that realize the modules described in \S\ref{sec:design:modules}: a query router to implement \groupModule, a metrics collector and recall estimator to implement the group monitor, and a controller to implement the \detectionModule and \mitigationModule.

\textbf{Query router.}
It implements the \groupModule and enforces per-{\cluster} search effort.
The \groupModule maps each incoming query to its corresponding {\cluster} by computing the nearest centroid, as described in \S~\ref{sec:design:modules}.
After determining the {\cluster}, the router retrieves the current search effort for that {\cluster} from a thread-safe dictionary maintained by the \rcheck controller.
It then injects the search effort value by issuing a PostgreSQL \texttt{SET LOCAL} command immediately before the vector search.
This approach requires no modifications to pgvector and only access to the standard PostgreSQL parameter interface.
All dictionary lookups and updates use atomic compare-and-swap operations to ensure queries never observe inconsistent state.

\textbf{Metrics collector.}
It implements the \groupMonitor and collects the statistics needed for the empirical tables.
We deploy a Prometheus~\cite{prometheus} instance alongside pgvector, coordinating both through Docker Compose.
At {\runtime}, the metrics collector records per-{\cluster} statistics: query count ($n_g$), throughput, latency, and current search effort ($e_g$).
These metrics are exposed via HTTP endpoints and scraped by Prometheus at a configurable \emph{monitoring interval}.
Shorter intervals enable faster adaptation to workload shifts but increase estimate variance and computation cost; longer intervals yield more stable measurements at the cost of slower response to recall differences from the target.
Because Prometheus pulls metrics asynchronously, the monitoring interval has no impact on query latency.

\textbf{Recall estimator.}
It implements the recall estimation functionality in the \groupMonitor.
We estimate per-{\cluster} recall by sampling \sample\% of queries per {\cluster} during each monitoring interval.
For each sampled query, a background thread executes an exact nearest-neighbor search (brute-force scan) and compares the result against the ANN result to compute recall.
These exact searches run in a dedicated thread pool and do not block the primary query path.
At the end of each monitoring interval, the estimator computes the per-{\cluster} recall estimate ($r_g$) as the mean recall across all samples and exports it to Prometheus alongside the other 
metrics.

\textbf{Controller.}
It implements the \detectionModule and \mitigationModule and maintains the empirical recall and performance tables ($T_r$ and $T_p$).
It runs as a dedicated process and coordinates the feedback loop at the end of each monitoring interval through three steps:

\emph{Step 1: Collect statistics.}
The controller queries Prometheus to retrieve per-{\cluster} metrics: query count, recall estimate, latency, and current search effort.

\emph{Step 2: Update tables and detect recall differences.}
It updates the empirical recall table $T_r$ and performance table $T_p$ with the new observations.
From these observations, the controller estimates how recall and latency change with search effort for each {\cluster} using linear regression.
The \detectionModule then identifies {\cluster}s where $|r_g - r^*| > \epsilon$, flagging them for intervention. 

\emph{Step 3: Compute and apply search effort adaptations.}
For each flagged {\cluster}, the \mitigationModule computes new search effort values using Algorithm~\ref{alg:rcheck_mitigation}.
Updated values are pushed to the query router via a single gRPC~\cite{indrasiri:book2020:grpc} call, which atomically refreshes the per-{\cluster} effort dictionary.

Our implementation decouples monitoring (Prometheus pulls metrics from the query router) from control (the controller pushes search effort adaptations), allowing each component to operate independently without blocking queries or introducing synchronization overhead.

\textbf{Parameters.}
We configure \rcheck with $k \in \{10, 50\}$ {\cluster}s depending on dataset size: $k=10$ for smaller datasets ({\fairface}, {\cifar}-100, {\dbpedia}) and $k=50$ for the larger dataset ({\glove}-100). 
We use a monitoring interval of 15 seconds. 
Search effort values follow ANN benchmarking defaults~\cite{aumuller2020ann}: \texttt{ef\_search} $\in [10, 500]$ for HNSW and \texttt{nprobe} $\in [1, 50]$ for IVF. 
Any discrete value within this range is valid.
The intervention thresholds, recall tolerance $\epsilon = 0.02$, safety margin $\lambda = 0.05$, and minimum throughput gain $\gamma = 5\%$, are set conservatively to reduce oscillation and filter out negligible adaptations.
We evaluate sensitivity to $k$ and the monitoring interval in \S\ref{sec:eval}.

\section{Experimental Methodology}
\label{sec:methodology}

\textbf{Experimental setup.}
We evaluate \rcheck using the widely-used pgvector\cite{pgvector}, a vector similarity search extension for
PostgreSQL16, as the underlying vector database~\cite{amazon-pgvector, aumuller2020ann, salesforce-pgvector, google-pgvector}. 
\rcheck is integrated directly into the
ANN Benchmarking framework~\cite{aumuller2020ann}, operating at the client layer without modifications to the database itself.

Our experiments run on a c6420 CloudLab node~\cite{cloudlab} with dual 16-core Intel Xeon CPUs (32 cores total) and 365GB of RAM. 
We allocate resources across containerized services deployed via Docker Compose. 
The pgvector service receives 32GB of memory and 8 CPU cores. 
\rcheck is deployed via Docker Compose alongside pgvector. 
Prometheus monitoring runs in a separate container with 1 CPU core and 256MB of memory; empirically, it uses less than 100MB of memory and $<$1\% CPU during typical 15-second monitoring intervals, so our allocation provides conservative headroom. 

\textbf{Datasets.}
We evaluate \rcheck on 
datasets spanning text, face recognition, vision, and knowledge-graph domains
\cite{pennington2014glove, karkkainen2021fairface, krizhevsky2009learning, auer2007dbpedia}.
For each dataset, we use a 90/10 train-test split: the training set (90\%) is indexed into the vector database, while the test set (10\%) composes the query workload.

(i)~\emph{\glove.} This is a text dataset made of 1.18M pre-trained word vectors (50–200 dimensions)~\cite{pennington2014glove}. 
These vectors were learned from a very large collection of public Twitter posts written in different languages~\cite{glovedatasets}. 

(ii)~\emph{\fairface.} It consists of 108{,}000 face images 
~\cite{karkkainen2021fairface}; 
we embed all images using the state-of-the-art InsightFace model~\cite{InsightFace}, producing 512-dimensional vector embeddings.

(iii)~\emph{\cifar-100.} It contains 60{,}000 images evenly distributed across 100 semantic classes~\cite{krizhevsky2009learning}; 
we generate 512-dimensional embeddings using widely-used CLIP ViT-B/32~\cite{radford2021learning}.

(iv)~\emph{\dbpedia.} It is derived from Wikipedia and represents structured knowledge-graph entities such as people, locations, and artworks \cite{auer2007dbpedia}. 
To avoid over-representation of highly popular entities, we construct a balanced subset by sampling page titles evenly across categories and alphabetic prefixes.
We embed each page title into a 384-dimensional vector using the popular \texttt{all-MiniLM-L6-v2} sentence embedding model \cite{all-MiniLM-L6-v2}, producing 53{,}000 vectors.

\textbf{Compared configurations.}
We evaluate \rcheck against pgvector and an \ideal. 

\emph{(1)~pgvector.}
We use pgvector as our baseline because it is one of the widely deployed open-source vector database extensions~\cite{amazon-pgvector, aumuller2020ann, salesforce-pgvector, google-pgvector}.
Pgvector supports two index types: {\hnsw}~\cite{HNSW} and {\ivfflat}~\cite{IVFFlat}.
We evaluate both to show that \rcheck generalizes across these index structures.
For each index, we sweep the global search effort parameter applied uniformly to all queries: \texttt{ef\_search} for {\hnsw} and \texttt{probes} for {\ivfflat}.
This sweep produces a curve characterizing the throughput-recall trade-off achievable with global tuning.
We highlight three representative operating points used throughout our evaluation:
\baselineLowEffort (low search effort, high throughput, lower recall);
\baselineBalancedEffort (medium search effort, balanced throughput and recall);
and \baselineHighEffort (high search effort, lower throughput, higher recall).

\emph{(2)}~\rcheck.
We evaluate \rcheck by sweeping the \emph{target recall} parameter.
This sweep produces a curve characterizing the throughput-recall trade-off achievable with {\cluster}-level adaptation.
We examine three representative operating points used throughout our evaluation:
\rcheckLtr (low target recall, high throughput, lower recall);
\rcheckMtr (medium target recall, balanced throughput and recall);
and \rcheckHtr (high target recall, lower throughput, higher recall).

\emph{(3)~\Ideal.}
We include an \ideal that represents the best achievable throughput-recall trade-off under per-query adaptation.
Given a target recall, the \ideal assigns each query its individually optimal search effort (i.e., the minimum value needed to achieve the target recall for that specific query).
This is unattainable in practice since it requires knowing each query's true nearest neighbors at \runtime.
However, it quantifies the maximum gains achievable through perfect per-query adaptation and serves as an upper bound for \rcheck.
We sweep the target recall to produce a curve comparable to pgvector and \rcheck.

\section{Evaluation}
\label{sec:eval}
We answer five questions.
\textbf{(Q1)} How does \rcheck reshape the throughput-recall trade-off compared to pgvector and \ideal, including effects on tail latency?
\textbf{(Q2)} How does the monitoring interval affect \rcheck's effectiveness?
\textbf{(Q3)} How does the number of clusters impact \rcheck's behavior?
\textbf{(Q4)} How robust is \rcheck to changes in ANN search parameters?
\textbf{(Q5)} What is \rcheck's overhead?
\subsection{Evaluating \rcheck's Effectiveness}
\label{subsec:q1}
\begin{figure}[t]
\centering 
\subfloat[][Mean recall vs. QPS.]{
  \includegraphics[clip,width=0.23\textwidth]{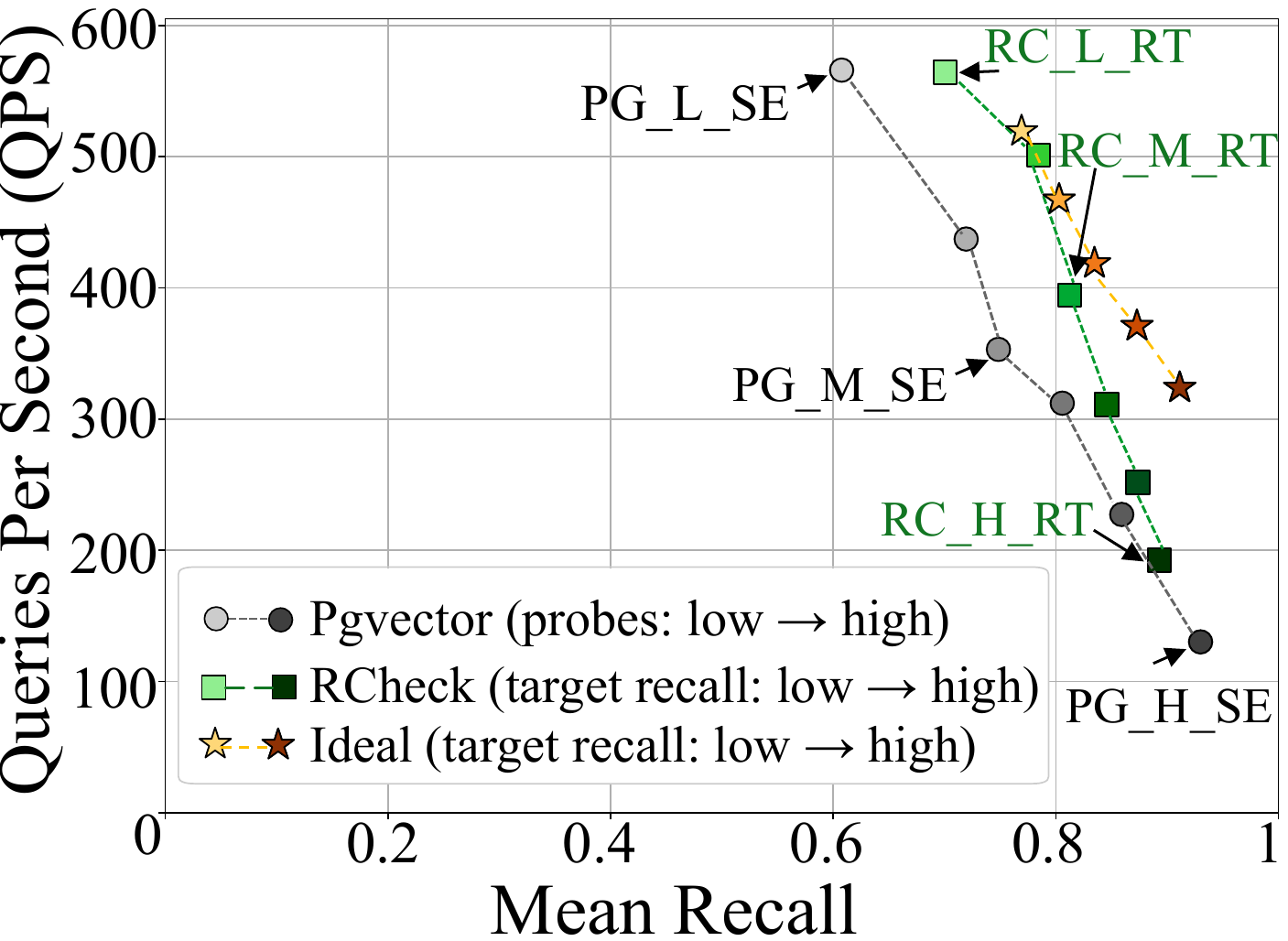}
  \label{fig:fig_glove100_mean_recall_vs_qps_overall_only}
}
\subfloat[][Recall distributions.]{
  \includegraphics[clip,width=0.23\textwidth]{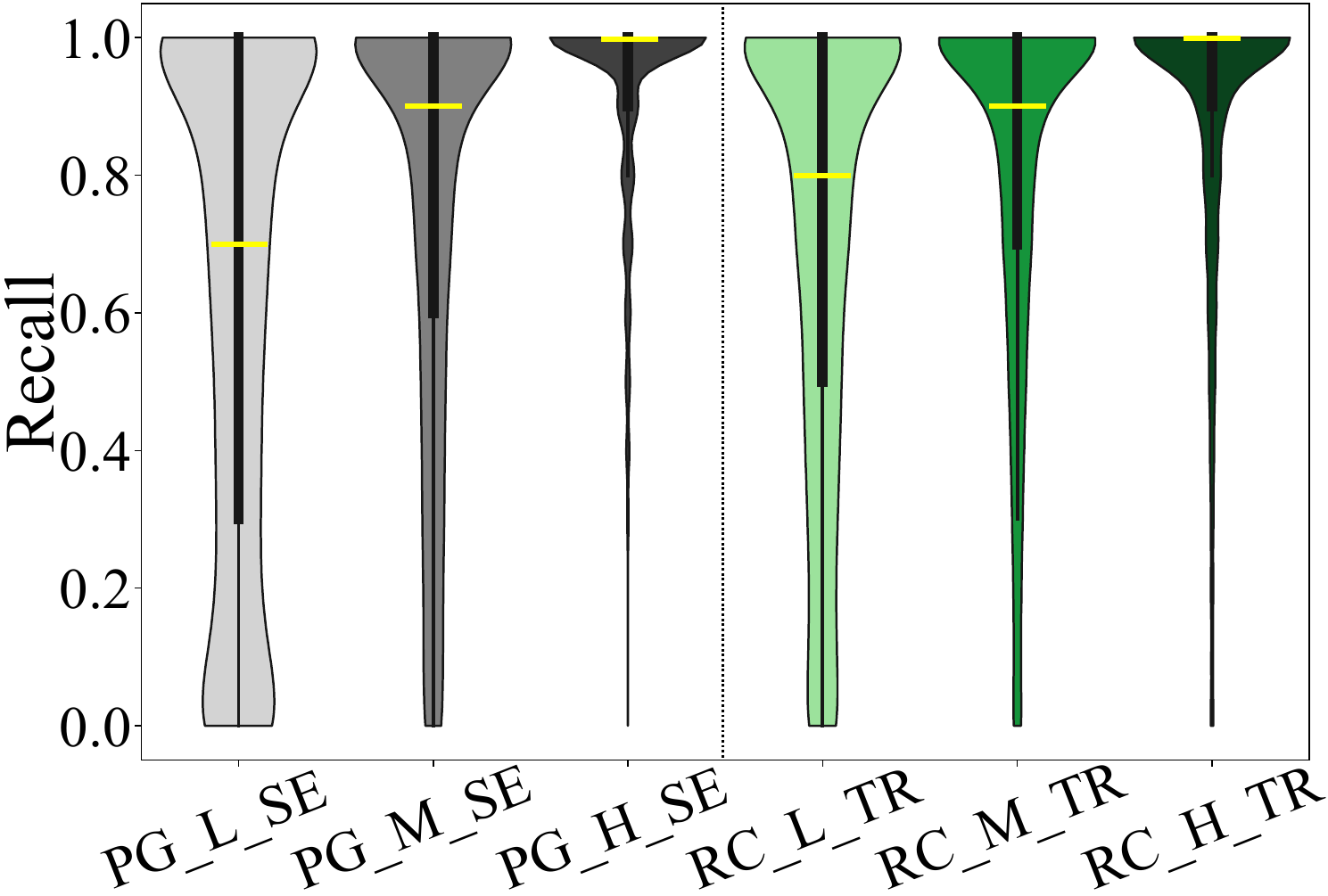}
  \label{fig:fig_glove100_overall_recall_violinplot_comparison_3comp}
}
\caption{
(a) QPS-recall trade-off comparison: \rcheck vs. pgvector vs. the ideal on {\glove}-100 with {\hnsw}. 
(b) Recall distributions across \rcheck and pgvector for selected 
points.
}
\vspace{-3mm}
\end{figure}
\subsubsection{{\glove}-100}
We run it on {\hnsw} with $k=50$ {\cluster}s.

\textbf{\QPSRecallAnalysis.}
Fig.~\ref{fig:fig_glove100_mean_recall_vs_qps_overall_only} shows QPS vs. mean recall for {\glove}-100 using an {\hnsw} index. 
Each curve represents the full range of achievable QPS vs. mean recall trade-offs for pgvector, \rcheck, and \ideal, obtained by sweeping corresponding tuning parameters, i.e., \texttt{ef\_search} for pgvector; target recall for \rcheck and ideal adaptation.

The \rcheck curve lies strictly above and to the right of the pgvector curve, achieving higher recall at any given throughput and higher throughput at any given recall.
For example, at target recall 0.85, \rcheck achieves 40.91\% higher throughput than pgvector. 
At 400 QPS, \rcheck improves recall from 0.73 to 0.81, a gain of 10.96\%.
These gains move \rcheck closer to the ideal curve.
The \ideal achieves 90.48\% higher throughput than pgvector at 0.85 recall. 
\rcheck closes 45\% of the gap to \ideal. The remaining gap exists because \rcheck adapts at the {\cluster} level rather than per query. Per-query adaptation would require tracking the impact of search effort on each query individually, necessitating ground truth computation for every query, a high cost. 
Nonetheless, \rcheck achieves 40.91\% higher throughput than pgvector, which does not adapt to reduce recall differences.

In deployment, operators do not need to sweep the target recall parameter. Instead, they 
specify their requirements: a target recall and a performance constraint (e.g., throughput).
\rcheck then automatically finds a search effort allocation across {\cluster}s that satisfies both. 
However, not all requirements are achievable. 
The \rcheck trade-off curve in Fig.~\ref{fig:fig_glove100_mean_recall_vs_qps_overall_only} determines what is achievable. 
If the requirements are infeasible, e.g., demanding 0.95 recall at 500 QPS when the curve shows 0.95 recall is only achievable at 200 QPS, \rcheck reports failure, and the operator must relax the recall target, the performance constraint, or both.

\textbf{\RecallDistAnalysis.}
Beyond mean recall, we examine how \rcheck reshapes the per-query recall distribution. 
Fig.~\ref{fig:fig_glove100_overall_recall_violinplot_comparison_3comp} compares pgvector and \rcheck at three pairs of operating points: (\baselineLowEffort vs. \rcheckLtr) and (\baselineBalancedEffort vs. \rcheckMtr) achieve similar throughput, while (\baselineHighEffort vs. \rcheckHtr) achieves similar mean recall, as described in~\S\ref{sec:methodology}.
To enable fair comparison across configurations that may use different internal target recalls, we evaluate all distributions against a fixed reference target of 0.85 ({\refTgtRecall}). 

We characterize each recall distribution using three metrics: (1) the fraction of queries meeting or exceeding {\refTgtRecall}, which measures how consistently the system delivers acceptable search quality~\cite{wang2024towards}; (2) the mean absolute error (MAE) relative to {\refTgtRecall};
and 
(3) the {expected shortfall} below {\refTgtRecall}, which is borrowed from risk analysis~\cite{rockafellar2000optimization, acerbi2002expected}, measures the average recall deficit among queries that fail to meet {\refTgtRecall}. 
 Together, these metrics quantify how often queries fail to meet {\refTgtRecall} (fraction below {\refTgtRecall}), how much queries typically deviate from the target (MAE), and how severe the worst-case failures are (shortfall).

The pgvector at \baselineLowEffort produces a wide distribution with 25th/75th percentiles of 0.3/1.0, spanning 0.7 recall points. 
Only 38.7\% of queries meet the {\refTgtRecall}, with MAE of 0.33 and shortfall of 0.29, indicating that queries missing the {\refTgtRecall} fall short by an average of 0.29 recall points. 
\rcheck at matched throughput (\rcheckLtr) tightens the 25th/75th percentiles to 0.5/1.0, reducing the spread to 0.5 recall points. 
The fraction of queries meeting {\refTgtRecall} increases to 46.7\%, MAE improves to 0.26, and shortfall decreases to 0.20. 
This 31\% reduction in shortfall means that when queries do fall short of the {\refTgtRecall}, they are closer to meeting it.

Moving to \baselineBalancedEffort, with pgvector, 55\% of queries meet  {\refTgtRecall}, with MAE of 0.35, shortfall of 0.17, and 25th/75th percentiles of 0.6/1.0. 
\rcheck at comparable throughput (\rcheckMtr) improves all three metrics: 62.6\% meet {\refTgtRecall}, MAE decreases to 0.18, shortfall drops to 0.11 and the distribution tightens the 25th/75th percentiles to 0.7/1.0.

We compare \baselineHighEffort and \rcheckHtr, which achieve similar mean recall, to evaluate how each distributes recall across queries.
With \baselineHighEffort, 81.8\% of queries meet {\refTgtRecall} (MAE: 0.15, shortfall: 0.04).
With \rcheckHtr at slightly lower mean recall, 78.3\% of queries meet {\refTgtRecall} (MAE: 0.16, shortfall: 0.06).
While \rcheckHtr's distributional metrics are marginally worse, it achieves 81.11\% higher throughput.
This shows that even at high recall targets, \rcheck delivers comparable search quality with substantially higher throughput.

{\textbf{\TailLatAnalysis.}} 
For \rcheckLtr vs. \baselineLowEffort, \rcheck faces slightly higher tail (p99) latency (3.11~ms vs. 2.23~ms) as it allocates additional search effort to low-recall {\cluster}s, while \baselineLowEffort applies minimal effort uniformly.
At higher recall targets, \rcheck significantly reduces tail latency. 
Compared to \baselineBalancedEffort, \rcheckMtr reduces p99 latency from 17.28~ms to 6.61~ms; compared to \baselineHighEffort, \rcheckHtr reduces p99 from 25.88~ms to 10.95~ms. 
Pgvector applies high search effort to all queries; \rcheck avoids unnecessary work on high-recall queries, reducing tail latency.

\subsubsection{{\fairface}} 
We run it on {\hnsw} with $k=10$ {\cluster}s.

\begin{figure}[t] \centering \subfloat[][Mean recall vs. QPS.]{ \includegraphics[clip,width=0.22\textwidth]{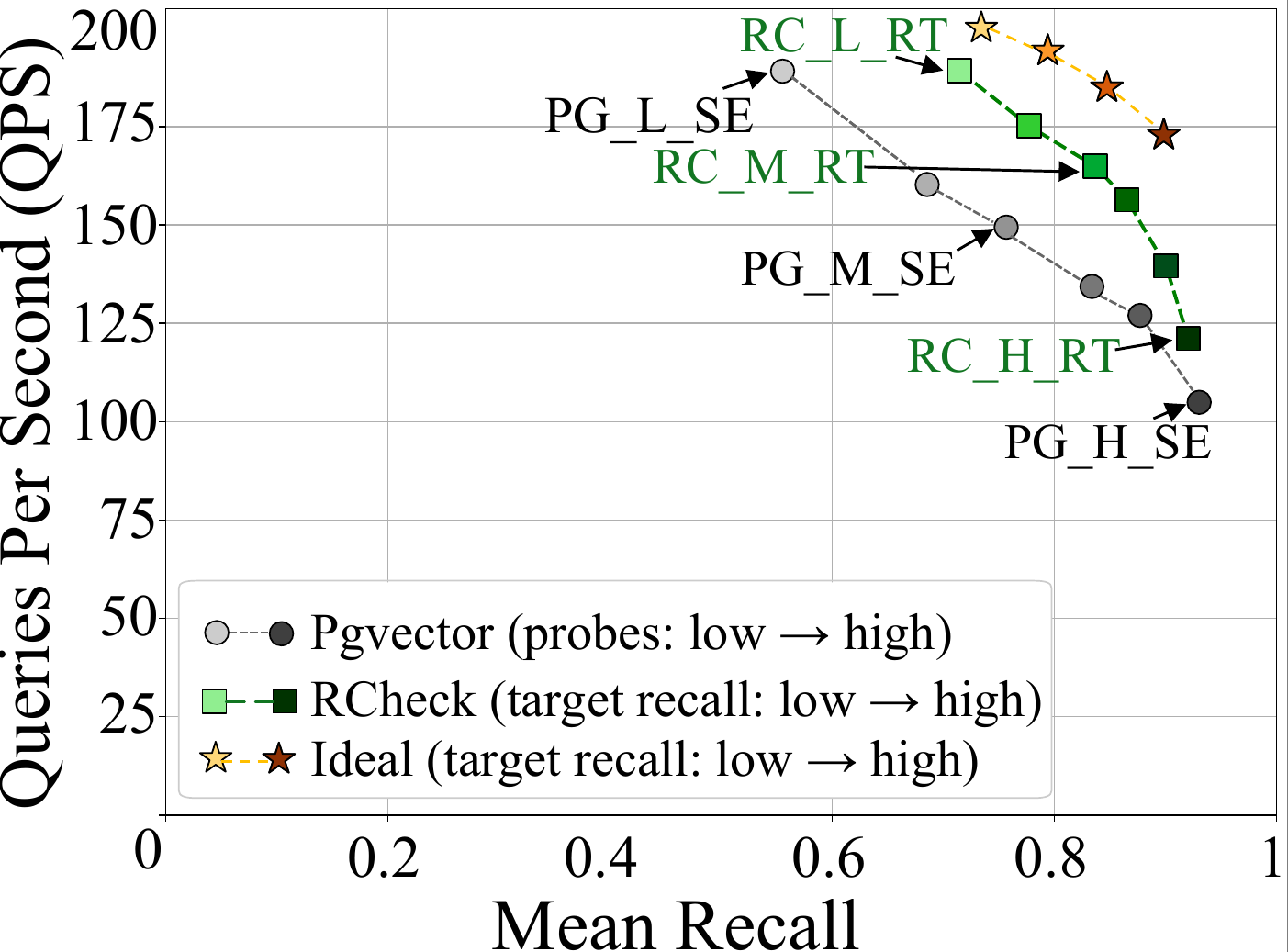} \label{fig:fig_fairface_mean_recall_vs_qps_overall_only} } \subfloat[][Recall distributions.]{ \includegraphics[clip,width=0.23\textwidth]{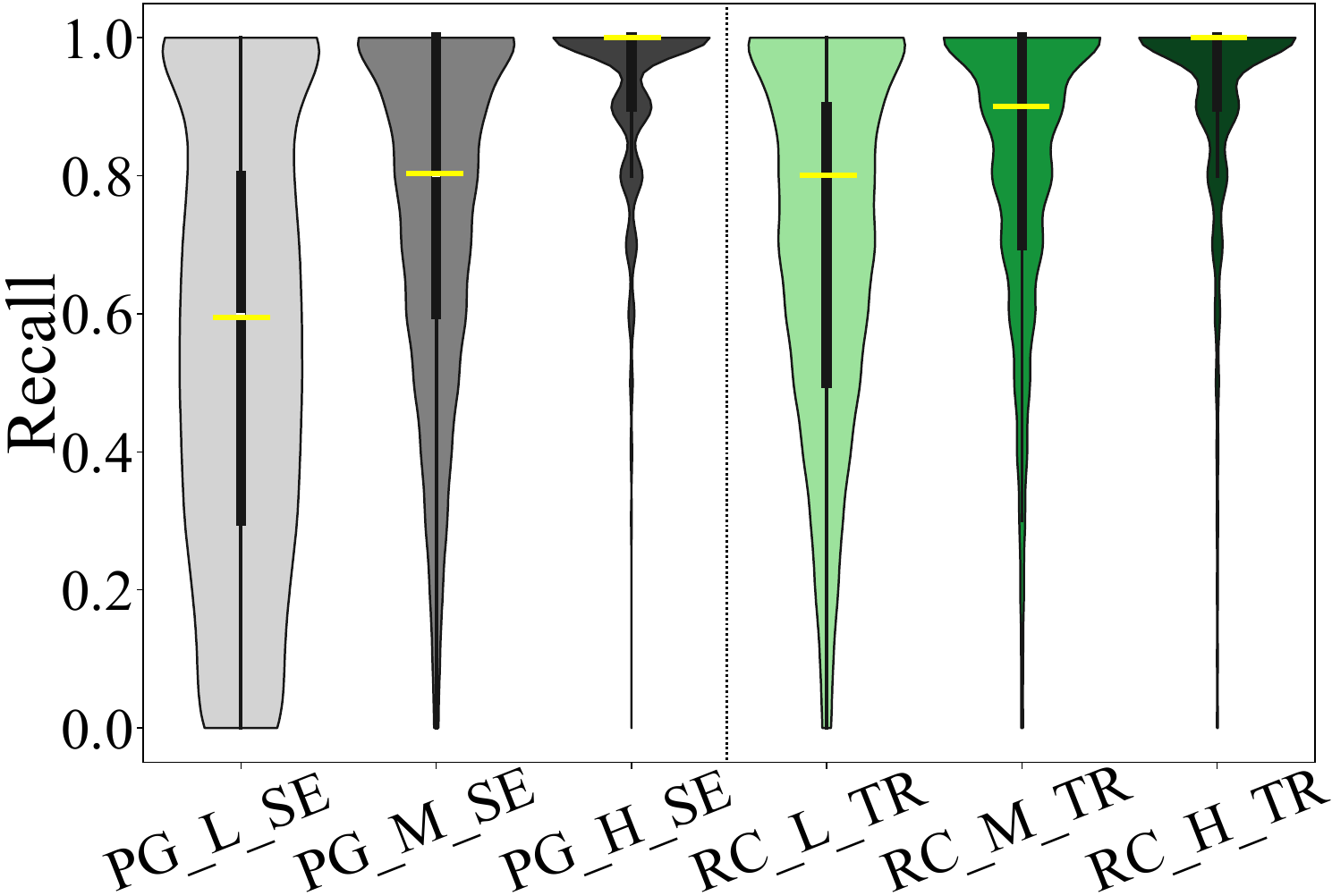} \label{fig:fig_fairface_overall_recall_violinplot_comparison_3comp} } \caption{ (a) QPS-recall trade-off comparison: \rcheck vs. pgvector vs. the ideal on {\fairface} with {\hnsw}. 
(b) Recall distributions across \rcheck and pgvector for selected points.
} \vspace{-3mm} \end{figure}

\textbf{\QPSRecallAnalysis.}
Fig.~\ref{fig:fig_fairface_mean_recall_vs_qps_overall_only} shows the QPS-recall trade-off, where \rcheck bridges the gap between pgvector and the \ideal.
At 0.85 recall, \rcheck achieves 25.63\% higher throughput than pgvector, closing 69.4\% of the gap to ideal.
At  187 QPS, \rcheck improves mean recall from 0.56 to 0.73, an improvement of 30.36\%.

\textbf{\RecallDistAnalysis.}
Fig.~\ref{fig:fig_fairface_overall_recall_violinplot_comparison_3comp} shows the recall distributions at three operating points. 
The pgvector at \baselineLowEffort achieves 24.3\% of queries meeting {\refTgtRecall}, with MAE of 0.35 and shortfall of 0.32 (25th/75th percentiles: 0.3/0.8). 
\rcheck at matched throughput (\rcheckLtr) improves to 38.2\% meeting target, MAE of 0.22, and shortfall of 0.17 (percentiles: 0.5/0.9). 
At \baselineBalancedEffort, the pgvector shows 45.1\% meeting target (MAE: 0.2, shortfall: 0.15, percentiles: 0.6/1), while \rcheck (\rcheckMtr) improves to 59.7\%, 0.15, and 0.08 (percentiles: 0.7/1). 
For \baselineHighEffort, the pgvector achieves 81.8\% meeting target (MAE: 0.13, shortfall: 0.03), while \rcheck (\rcheckHtr) at comparable mean recall shows 80.3\% meeting target, MAE of 0.14, and shortfall of 0.04, with 19.41\% higher throughput.

{\textbf{\TailLatAnalysis.}} 
Tail latencies follow a similar trend as {\glove}-100: for example,  \rcheckMtr vs. \baselineBalancedEffort, \rcheck reduces p99 from 24.13 ms to 7.29 ms. 
\subsubsection{{\cifar}-100}
We run it on {\ivfflat} with $k=10$ {\cluster}s to evaluate \rcheck on a different index type.

\begin{figure}[t]
\centering 
\subfloat[][Mean recall vs. QPS.]{
  \includegraphics[clip,width=0.22\textwidth]{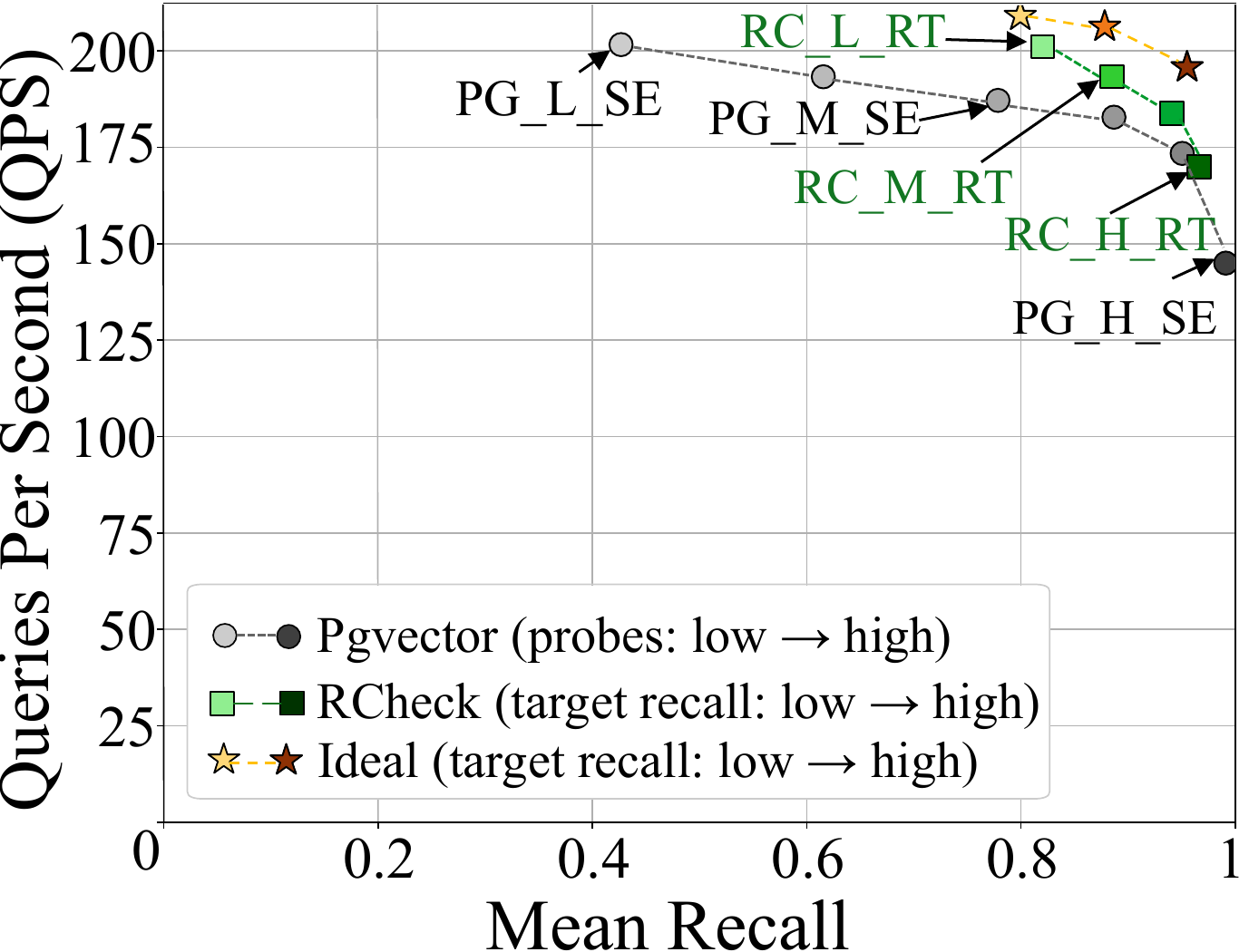}
  \label{fig:fig_cifar_mean_recall_vs_qps_overall_only}
}
\subfloat[][Recall distributions.]{
  \includegraphics[clip,width=0.23\textwidth]{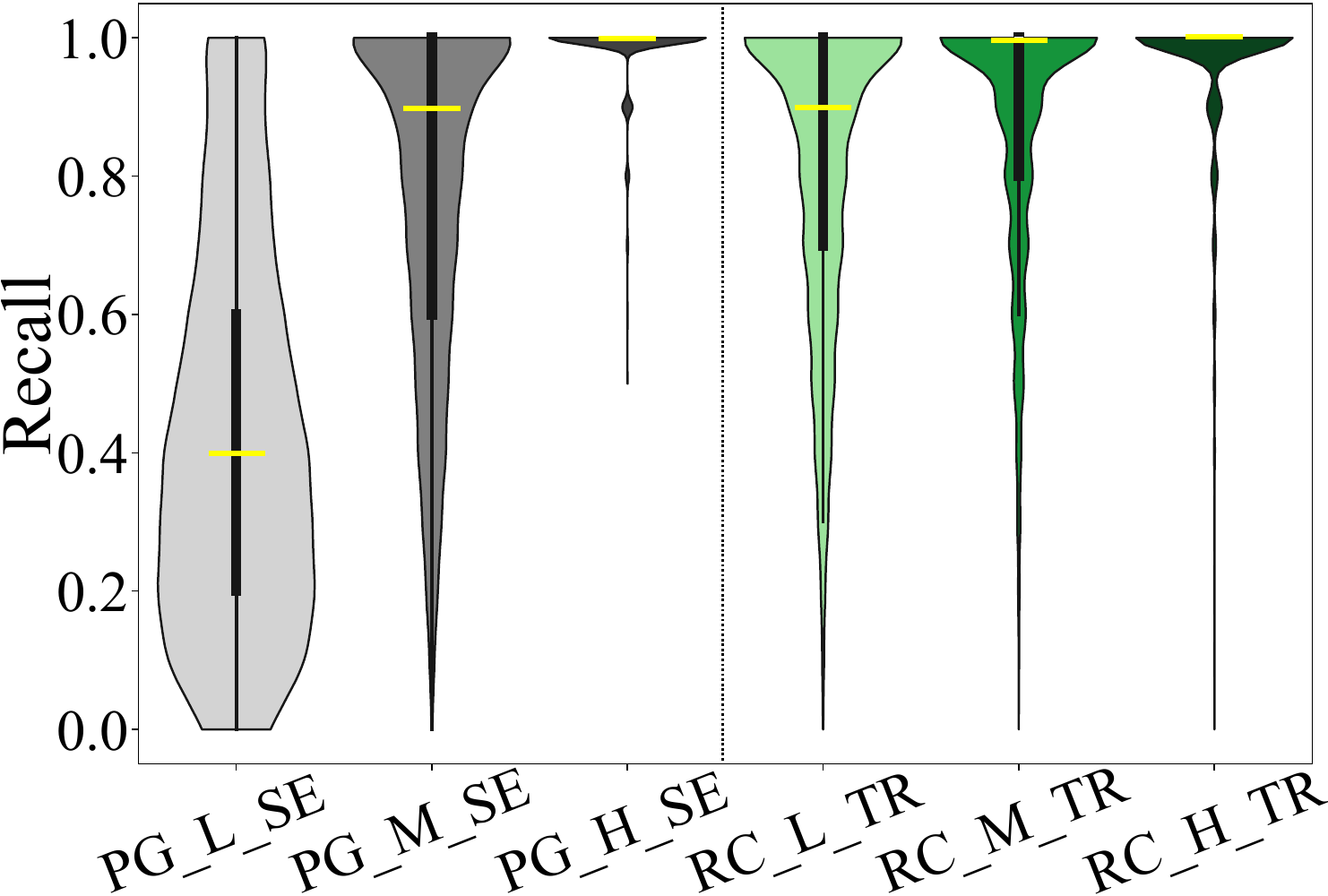}
  \label{fig:fig_cifar_overall_recall_violinplot_comparison_3comp}
}
\caption{
(a) QPS-recall trade-off comparison: \rcheck vs. pgvector vs. the ideal on {\cifar}-100 with {\ivfflat}. 
(b) Recall distributions across \rcheck and pgvector for selected points.
}
\vspace{-3mm}
\end{figure}

\textbf{\QPSRecallAnalysis.}
Fig.~\ref{fig:fig_cifar_mean_recall_vs_qps_overall_only} 
shows that \rcheck's benefits extend beyond {\hnsw}. 
At 0.8 recall, \rcheck achieves 6.95\% higher throughput than pgvector, closing 59.1\% of the gap to \ideal. 
At 200 QPS, \rcheck improves recall from 0.42 to 0.81, a 92.86\% gain.

\textbf{\RecallDistAnalysis}
The recall distribution improvements follow the same pattern as before.
At \baselineLowEffort, the pgvector achieves 10.66\% meeting {\refTgtRecall}, with MAE of 0.44 and shortfall of 0.43 (percentiles: 0.2/0.6), while \rcheck (\rcheckLtr) improves to 57.91\% meeting target, MAE of 0.17, and shortfall of 0.1 (percentiles: 0.7/1). 
At \baselineBalancedEffort, the pgvector shows 50.8\% meeting target, with MAE of 0.2 and shortfall of 0.13 (percentiles: 0.6/1), while \rcheck (\rcheckMtr) achieves 71.8\% meeting target, MAE of 0.15, and shortfall of 0.05 (percentiles: 0.8/1). 
For \baselineHighEffort, the pgvector reaches 97.1\% meeting target, with MAE of 0.14 and shortfall of 0, while \rcheck (\rcheckHtr) shows 92.3\% meeting target, MAE of 0.14, and shortfall of 0.01, with 16.89\% higher throughput.

{\textbf{\TailLatAnalysis.}} 
Tail latencies follow a similar trend as prior (e.g., 6.52~ms vs. 25.61~ms for \rcheckMtr vs. \baselineBalancedEffort).

\subsubsection{{\dbpedia}}
 We run it on {\ivfflat} with $k=10$ {\cluster}s.

\begin{figure}[t]
\centering 
\subfloat[][Mean recall vs. QPS.]{
  \includegraphics[clip,width=0.23\textwidth]{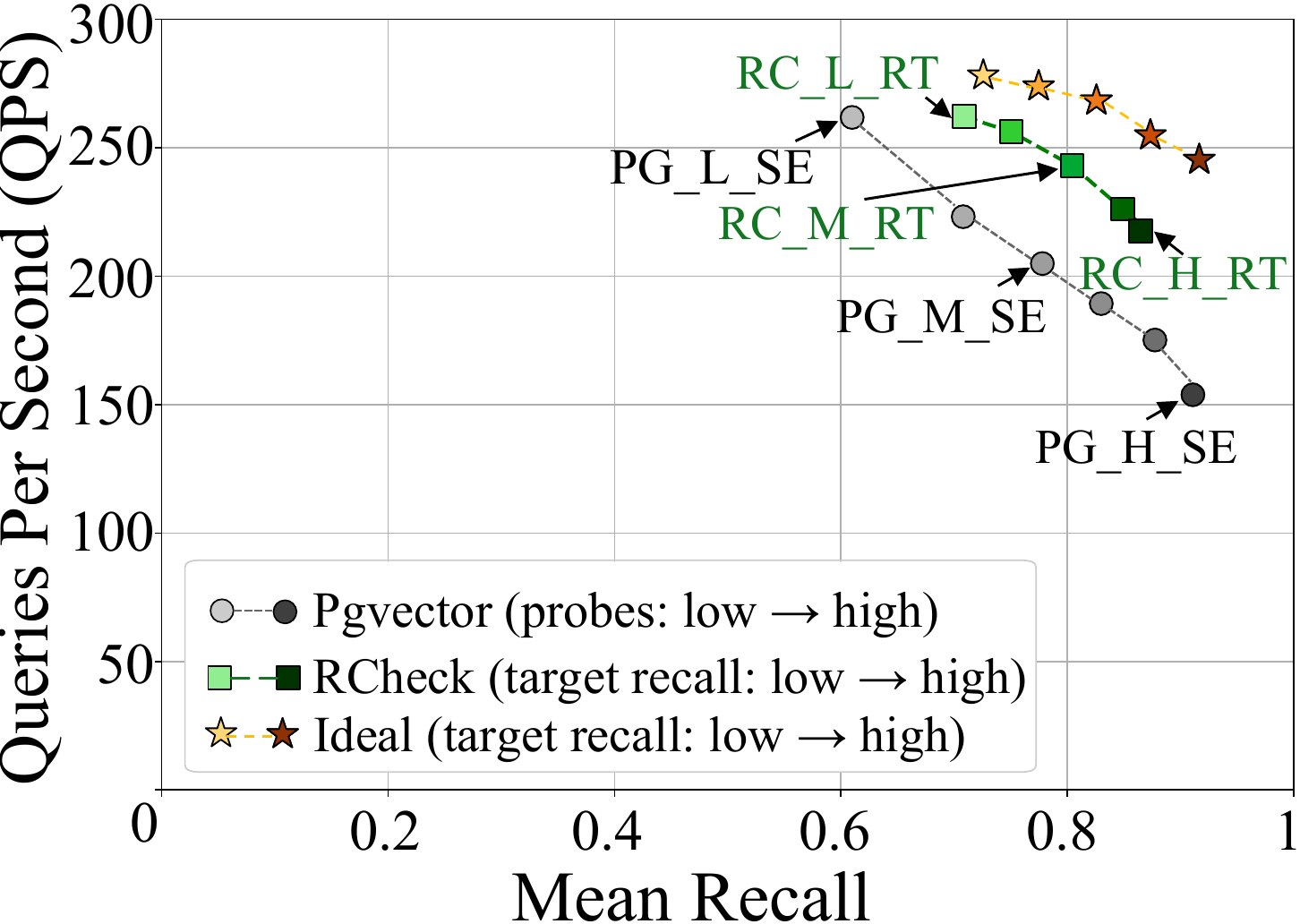}
  \label{fig:fig_dbpedia_mean_recall_vs_qps_overall_only}
}
\subfloat[][Recall distributions.]{
  \includegraphics[clip,width=0.23\textwidth]{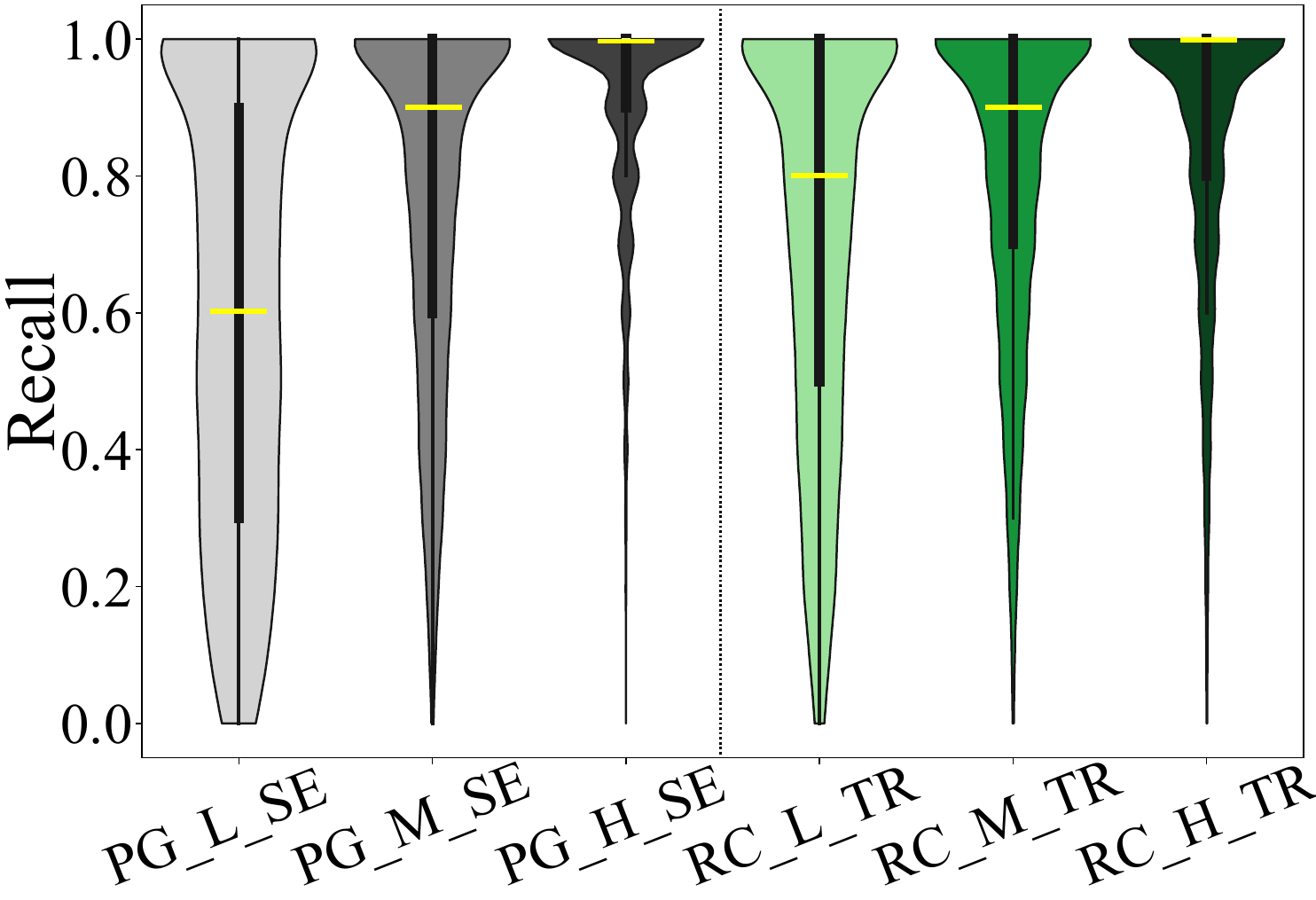}
  \label{fig:fig_dbpedia_overall_recall_violinplot_comparison_3comp}
}
\caption{
(a) QPS-recall trade-off comparison: \rcheck vs. pgvector vs. the ideal on {\dbpedia} with {\ivfflat}. 
(b) Recall distributions across \rcheck and pgvector for selected points.
}
\vspace{-3mm}
\end{figure}

\textbf{{\QPSRecallAnalysis}.}
Fig.~\ref{fig:fig_dbpedia_mean_recall_vs_qps_overall_only} shows that \rcheck maintains its advantage across the operating range, achieving 23.6\% higher throughput than pgvector at 0.85 target recall and covering 51.2\% of the gap to ideal.
At 225 QPS, \rcheck improves mean recall from 0.71 to 0.84, a gain of 20\%.

\textbf{\RecallDistAnalysis.}
The recall distribution analysis shown in Fig.~\ref{fig:fig_dbpedia_overall_recall_violinplot_comparison_3comp} follows the prior pattern. 
At \baselineLowEffort, the pgvector achieves 31.73\% meeting {\refTgtRecall} (MAE: 0.31, shortfall: 0.27, percentiles: 0.3/0.9), while \rcheck (\rcheckLtr) improves to 42.6\%, 0.24, and 0.19 (percentiles: 0.5/1). 
At \baselineHighEffort, the pgvector reaches 77\% (MAE: 0.14, shortfall: 0.03), while \rcheck (\rcheckHtr) shows 69.8\%, 0.15, and 0.07, with 41.33\% higher throughput.

\textbf{\TailLatAnalysis.}
Tail latency trends hold: \rcheckMtr achieves 5.12~ms vs. 23.5~ms for \baselineBalancedEffort.

\begin{figure}[t]
\centering 
\subfloat[][Varying the monitoring interval]{
  \includegraphics[clip,width=0.23\textwidth]{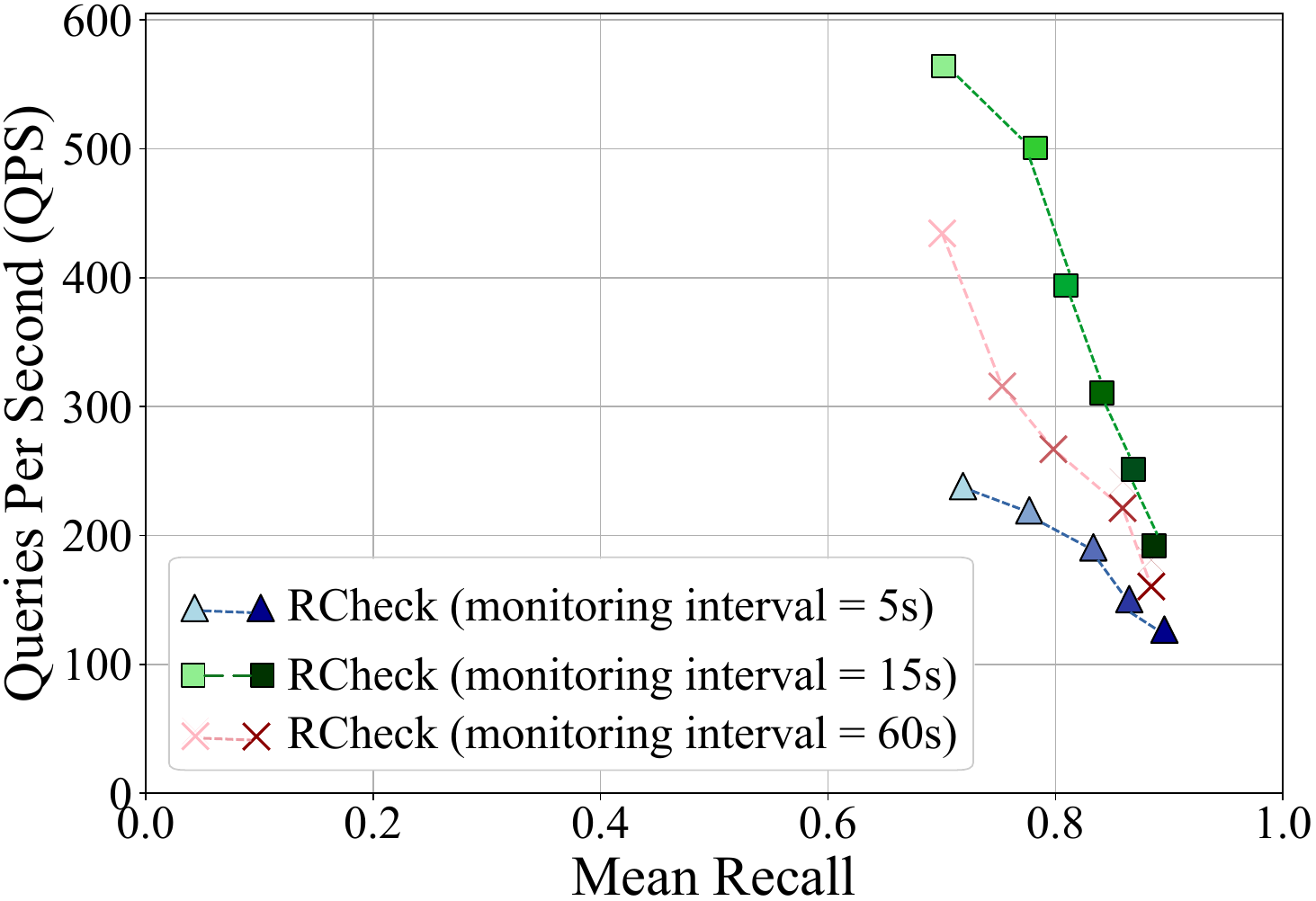}
  \label{fig:fig_glove100_monitoring_interval_sens_mean_recall_vs_qps_overall_only}
}
\subfloat[][Varying the number of {\cluster}s]{
  \includegraphics[clip,width=0.23\textwidth]{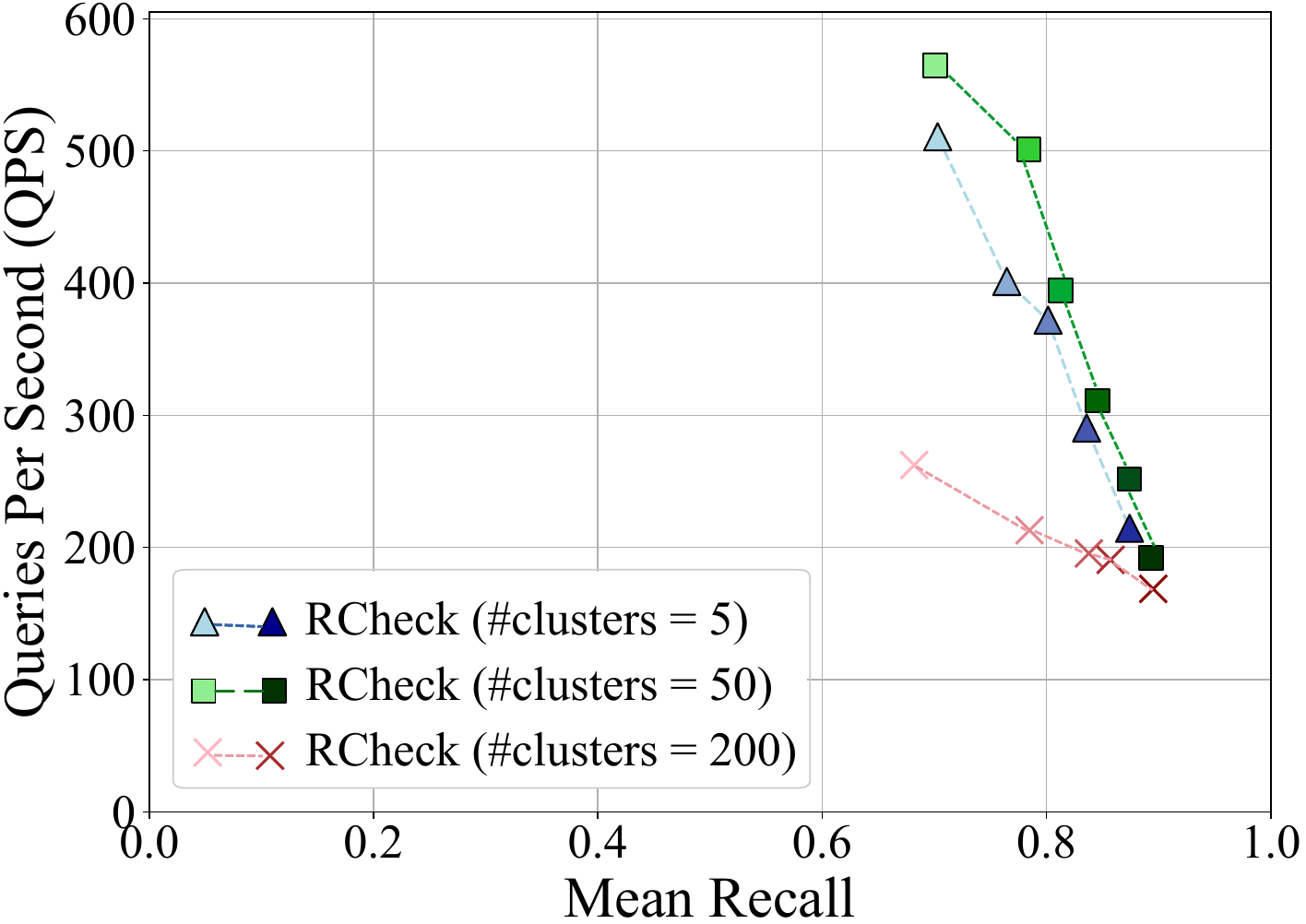}
  \label{fig:fig_glove100_cluster_number_sens_mean_recall_vs_qps_overall_only}
}
\caption{
QPS-recall trade-off on {\glove}-100 with {\hnsw}, showing \rcheck's sensitivity to (a) monitoring interval length and (b) number of {\cluster}s.
}
\vspace{-3mm}
\end{figure}
\subsection{{Evaluating the Monitoring Interval's Impact}}
We study how the monitoring interval affects \rcheck by varying it between 5s, 15s, and 60s, shown in Fig.~\ref{fig:fig_glove100_monitoring_interval_sens_mean_recall_vs_qps_overall_only}.
A short interval (5s) reacts quickly but relies on limited samples, leading to noisy recall estimates and frequent adaptations. 
At 0.85 target recall, this overhead reduces throughput to 190 QPS compared to 310 QPS for 15s configuration, which is a 38.7\% loss. 
At matched throughput of 240 QPS, mean recall drops to 0.72, underperforming 15s by 17.24\%.

A long interval (60s) produces more stable estimates and reduces adaptation overhead, achieving 250 QPS at 0.85 target recall, which is better than 5s but still 19.35\% below 15s. 
However, low-recall {\cluster}s persist for up to 60s before adaptation, causing recall degradation: at 310 QPS, mean recall is 0.75, 11.76\% below 15s.

The moderate interval (15s) balances stability and responsiveness. It achieves the best recall-throughput trade-off, 
operating closest to the ideal oracle across the entire range.
\subsection{{\Cluster} Granularity Impact on \emph{\rcheck}}
We evaluate how the number of {\cluster}s affects \rcheck by varying {\cluster} granularity while keeping the index and dataset fixed (Fig.~\ref{fig:fig_glove100_cluster_number_sens_mean_recall_vs_qps_overall_only}).
With few {\cluster}s (5), each {\cluster} covers a large region of the embedding space, grouping queries with different recall behavior together. 
This limits \rcheck's ability to identify and correct localized recall problems: 
At the 0.85 target recall, this configuration achieves 280 QPS, which is a 10.71\% throughput loss compared to 50 {\cluster}s.
At matched throughput (500 QPS), mean recall drops to 0.71, 10.13\% below the 50-{\cluster} configuration.

With many {\cluster}s (200), each {\cluster} becomes too small and receives too few queries for reliable recall estimation.
This leads to frequent, noisy adaptations that increase overhead.
At the 0.85 target recall, this configuration achieves only 200 QPS, which is 35.48\% lower than 50 {\cluster}s.
At matched throughput (210 QPS), mean recall is 0.78, which is 11.36\% worse than even the 5-{\cluster} configuration.

Fig.~\ref{fig:fig_glove100_cluster_number_sens_mean_recall_vs_qps_overall_only} shows that 50 {\cluster}s achieves the best balance.
At this number of groups, {\cluster}s are specific enough to isolate low-recall regions while receiving sufficient queries for stable recall estimation.

\begin{figure}[t]
\centering 
\subfloat[][Varying embedding model.]{
  \includegraphics[clip,width=0.23\textwidth]{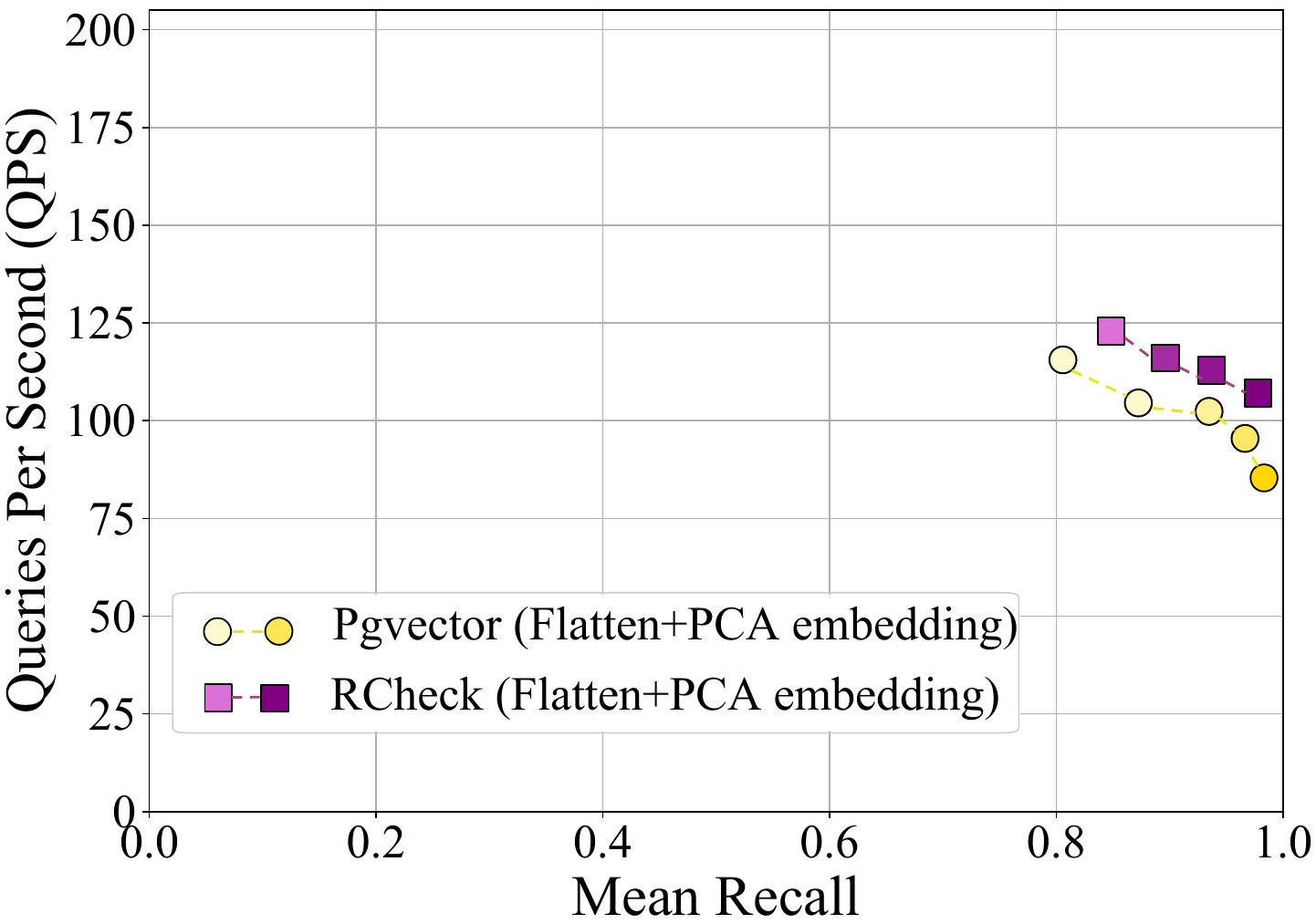}
  \label{fig:fig_fairface_embedding_sens}
}
\subfloat[][Varying \texttt{M} in \hnsw.]{
  \includegraphics[clip,width=0.23\textwidth]{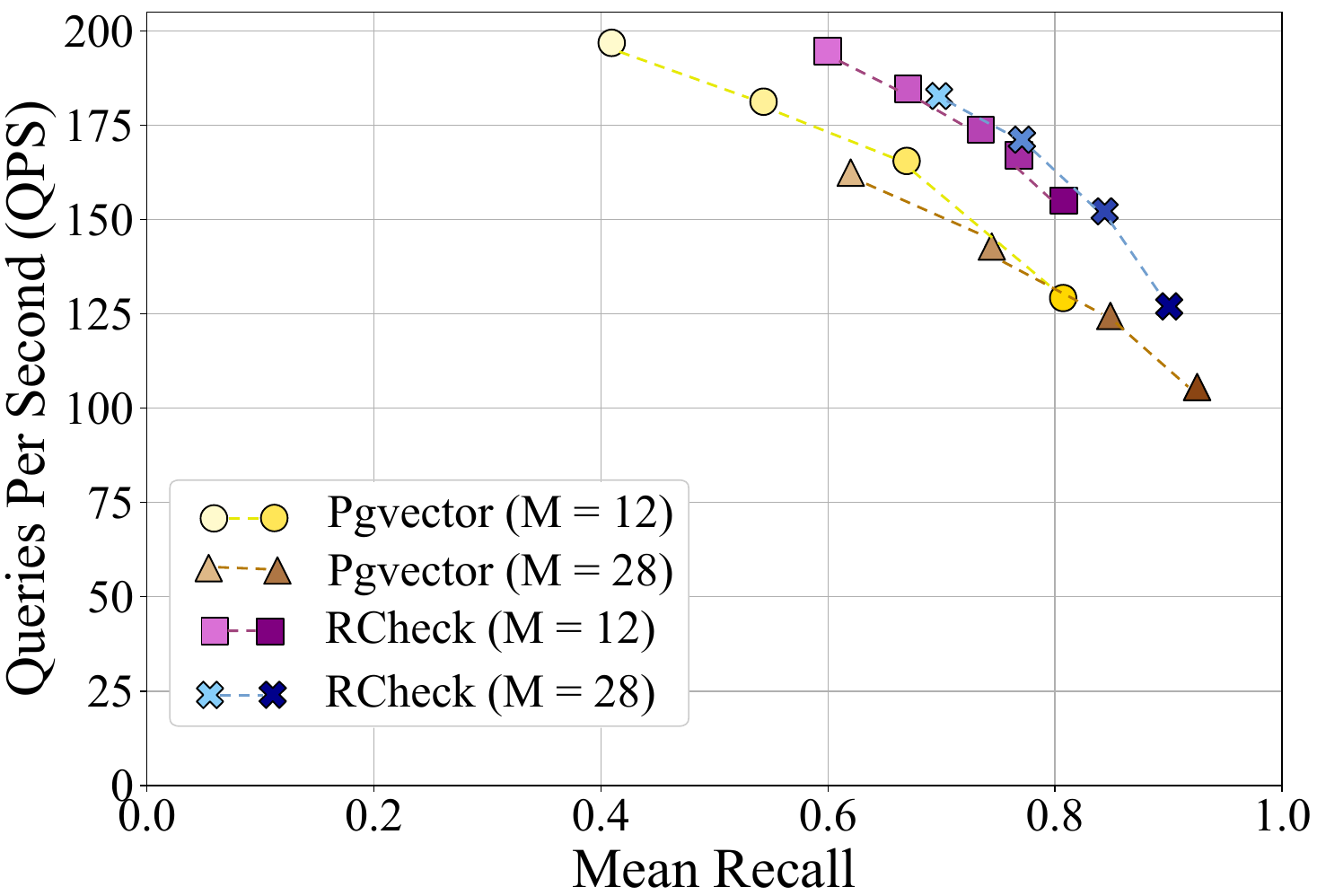}
  \label{fig:fig_fairface_M_sens}
}
\caption{
\rcheck improves the QPS-recall trade-off on {\hnsw} for the {\fairface} dataset (a) when changing the embedding model and (b) across different \texttt{M} values.
}
\vspace{-3mm}
\end{figure}

\begin{figure}[t]
\centering 
\subfloat[][Varying \texttt{ef\_construction}.]{
  \includegraphics[clip,width=0.23\textwidth]{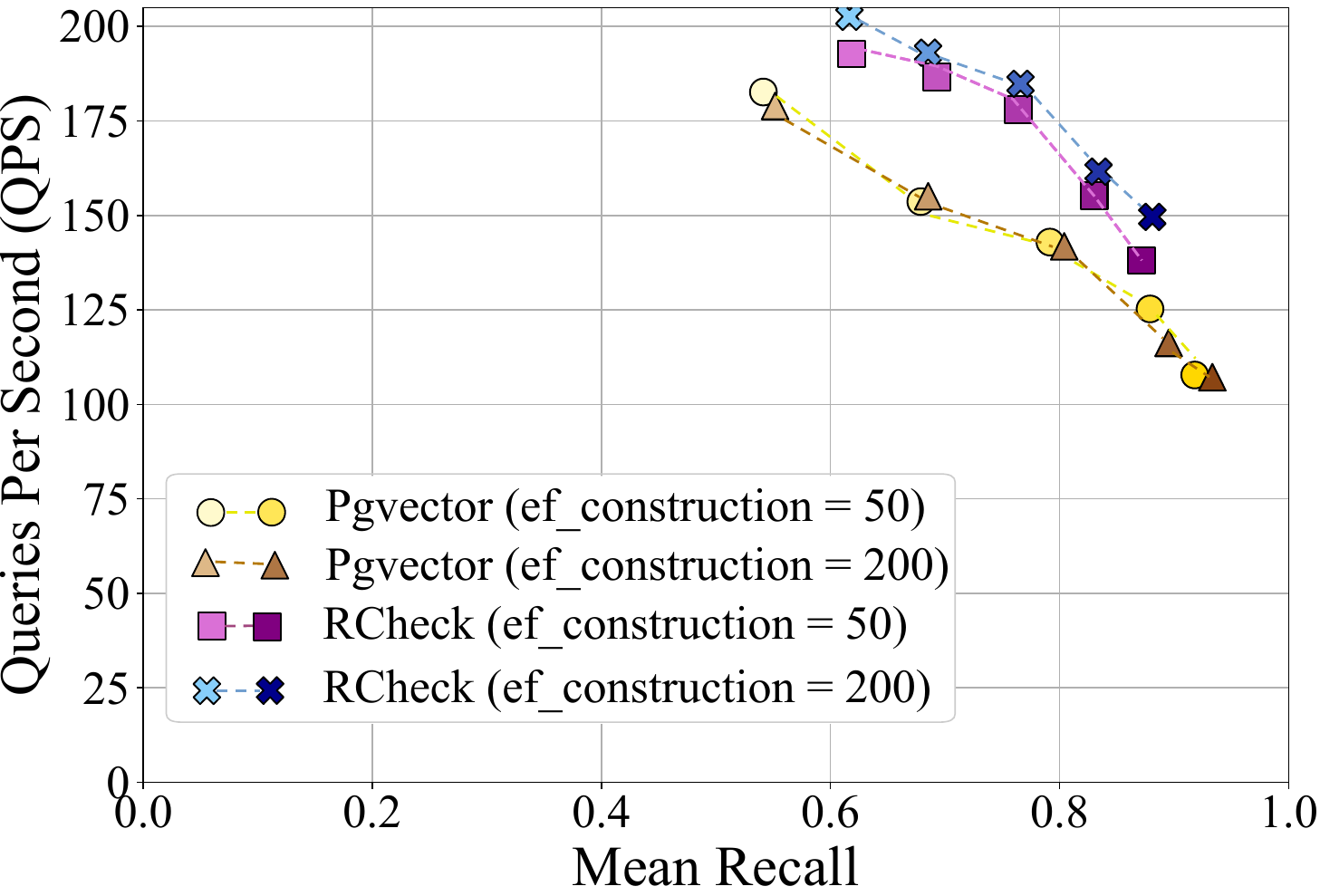}
  \label{fig:fig_fairface_ef_construction_sens}
}
\subfloat[][Varying \texttt{nlist} in \ivfflat.]{
  \includegraphics[clip,width=0.23\textwidth]{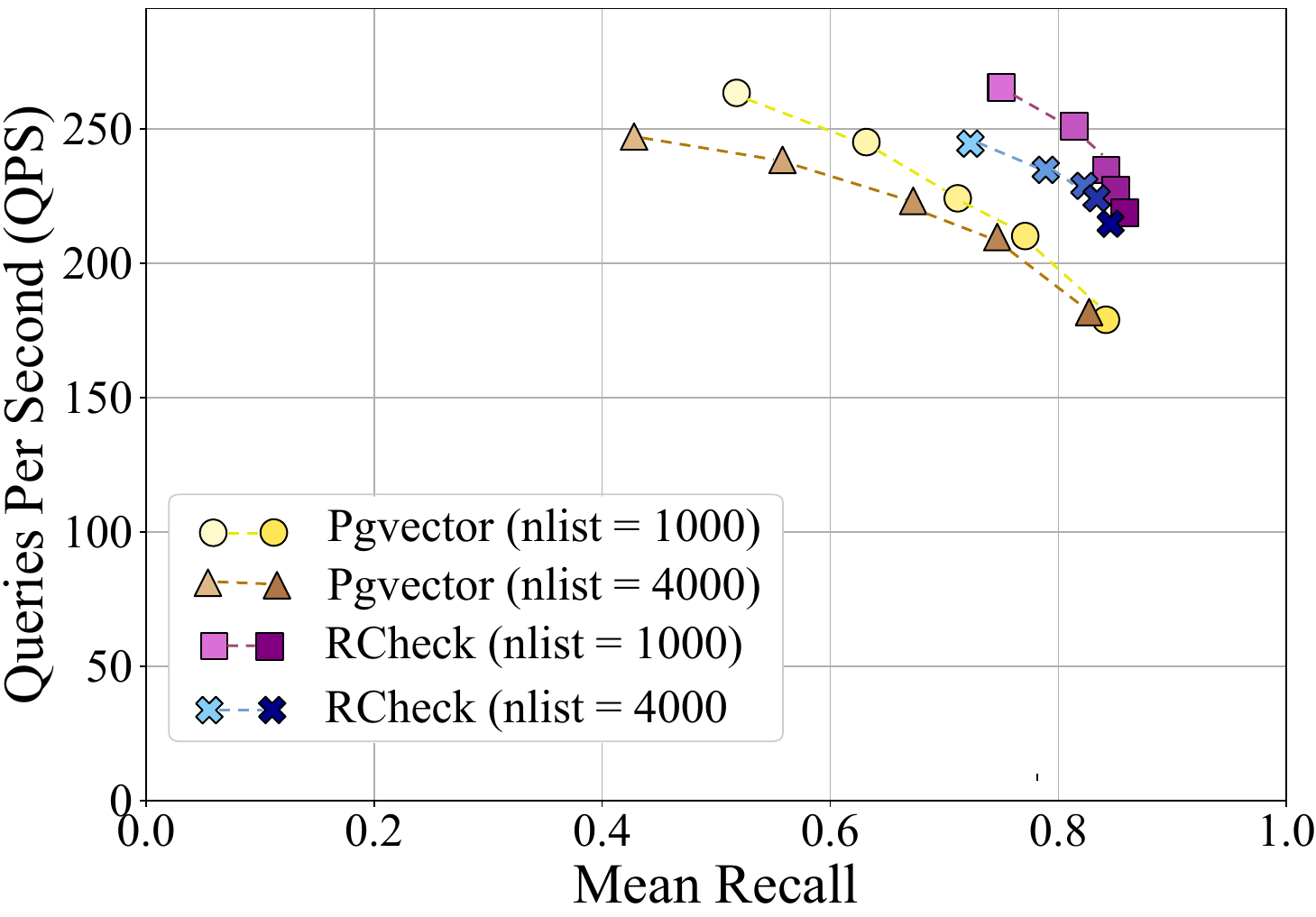}
  \label{fig:fig_dbpedia_nlist_sens}
}
\caption{
\rcheck improves the QPS-recall trade-off on (a)~{\hnsw} for the {\fairface} dataset across different \texttt{ef\_construction} values and (b)~{\ivfflat} for the {\dbpedia} dataset across different \texttt{nlist} values.
}
\vspace{-3mm}
\end{figure}
\vspace{-1em}
`\subsection{\emph{\rcheck}'s Robustness to ANNS Parameters}
We evaluate \rcheck's robustness to embedding model and index construction parameters (\S\ref{sec:motivation}). 

\textbf{Embedding model variation.}
To assess sensitivity to embedding model choice, we repeat our experiments with alternative embeddings for {\fairface}: we flatten the original vectors and apply Principal Component Analysis (PCA)~\cite{abdi2010principal}\todo{cite this} to obtain 1024-dimensional embeddings.
Fig.~\ref{fig:fig_fairface_embedding_sens} shows that \rcheck maintains its advantage: at 0.85 target recall, \rcheck achieves 125 QPS compared to pgvector's 107 QPS (16.82\% improvement), consistent with the behavior observed with the original embeddings.

\textbf{HNSW construction parameters.}
For {\hnsw} on {\fairface}, we evaluate robustness by varying key construction parameters from values used in \S\ref{subsec:q1}.
Our previous experiments used \texttt{M}=24; here we test with \texttt{M}=12 and \texttt{M}=28, shown in Fig.~\ref{fig:fig_fairface_M_sens}.
With \texttt{M}=12, \rcheck achieves 153 QPS at 0.8 recall vs.\ pgvector's 126 QPS, which is a 21.4\% improvement.
With \texttt{M}=28, \rcheck achieves 151 QPS at 0.85 recall vs.\ pgvector's 125 QPS, a 20.8\% improvement.

Similarly, our previous {\fairface} experiment in \S\ref{subsec:q1} used \texttt{ef\_construction}=100; here we test with \texttt{ef\_construction}=50 and \texttt{ef\_construction}=200, as shown in Fig.~\ref{fig:fig_fairface_ef_construction_sens}.
Across all parameter settings, \rcheck consistently improves throughput by 20-40\% at the target recall.

For {\ivfflat} ({\dbpedia}), we similarly vary \texttt{nlist} by halving and doubling its default value as shown in ~\ref{fig:fig_dbpedia_nlist_sens}. 
\rcheck again preserves its advantage across these settings, improving the trade-off between recall and throughput.
\vspace{-1em}
\subsection{Evaluating \rcheck's Overhead}
\rcheck introduces the following overheads.
{\Cluster} creation is a one-time cost during setup; for $k \in \{5, 50\}$, it takes 260~seconds on average across datasets, adding $\sim$30\% overhead relative to index build time on average, and runs in parallel with index construction.
Assigning each query to its {\cluster} takes $<$0.1~ms, negligible compared to typical query latencies of 3--9~ms in our experiments.
Recall 
estimation samples 2\% of queries per {\cluster} in each monitoring interval to estimate {\cluster}-level recall, running in parallel with query processing and consuming $<$50\% of CPU cycles.
\Runtime adaptation computes new search effort allocations in $<$14~ms, which is $<$0.1\% of the 15-second monitoring interval and runs asynchronously without blocking queries.
\section{Discussion}
\label{sec:discussion}
\textbf{{\Runtime} adaptation under performance constraints.}
We design \rcheck to operate strictly at {\runtime} 
without modifying the embedding model or the ANN index structure. 
This design choice avoids expensive index rebuilds and allows the system to react quickly to recall degradation under changing workloads.
However, it also limits what \rcheck can correct. 
\rcheck assumes that the target recall is achievable within the available performance budget. 
If the target recall is unachievable because the index quality is too low or the performance budget is too tight, no amount of search effort tuning will suffice.
\rcheck detects this condition when {\cluster}s persistently fall short of the target despite operating at maximum allowed search effort.
In such cases, the system alerts the operator that the current recall target is unachievable within the performance budget, requiring offline intervention or revised constraints.


\textbf{Search effort reallocation.}
A key insight behind \rcheck is that {\cluster}s exceeding the recall target can reduce their search effort to create computation headroom for {\cluster}s falling short.
This introduces coupling between {\cluster}s: improving a {\cluster}'s recall may require reducing effort elsewhere.
Our current design prioritizes {\cluster}s by recall difference magnitude from the target, allocating more effort to those furthest below the target.
This aligns with reducing maximum difference from the target recall.
More expressive policies like protecting certain {\cluster}s from effort reduction could be supported, but might need explicit operator input.


\section{Related Work}
\label{sec:related_work}

\textbf{Global parameter tuning.}
ANNS has been extensively studied, with widely deployed indexes including graph-based methods such as HNSW~\cite{HNSW, malkov2018efficient} and DiskANN~\cite{jayaram2019diskann}, and partition-based approaches such as {\ivfflat}~\cite{IVFFlat}, IVF~\cite{jegou2010product}, and FAISS~\cite{douze2025faiss}.
These systems expose search effort parameters (e.g., \texttt{ef\_search}) that trade recall for throughput.
In practice, operators select global parameter values via benchmarking frameworks like ANN-Benchmarks~\cite{aumuller2020ann} or vendor recommendations~\cite{douze2025faiss, pgvector}, applying a single configuration to all queries.
This approach assumes uniform recall behavior across queries and does not account for per-query or per-region variation in search quality.
In contrast, \rcheck operates at {\runtime} and adapts search effort at a finer granularity based on observed recall.

\textbf{Adaptive index structures}
Prior work has addressed workload heterogeneity by adapting ANN index structures. 
SPANN~\cite{chen2021spann} identifies the closest clusters in memory and then reads only the necessary data blocks from the disk to find the nearest neighbors. 
This query-aware approach ensures the system only performs minimal disk reads, keeping search speeds fast. 
QUAKE~\cite{mohoney2025quake} employs a multi-level partitioning scheme that reorganizes partitions in response to updates and changing access patterns, guided by a cost model predicting query latency.
These approaches modify the index structure, requiring maintenance overhead and potential query disruption during reorganization.
In contrast, \rcheck operates at \runtime by adopting search effort parameters without modifying the underlying index, enabling immediate adaptation.

\textbf{Differences in search quality.}
Recent work has examined fairness in information retrieval, focusing on biases in ranked outputs~\cite{singh2018fairness, zehlike2017fa, castillo2019fairness} and in embedding representations~\cite{caliskan2017semantics, bolukbasi2016man, rekabsaz2021societal}.
For example, Singh et al.~\cite{singh2018fairness} propose exposure fairness for rankings, while Zehlike et al.~\cite{zehlike2017fa} study group fairness in top-$k$ results.
These approaches primarily operate at the ranking or representation level.
More recently, Wang et al.~\cite{wang2024towards} also questioned the adequacy of mean recall as a system-level metric and introduced \emph{robustness} as a metric that captures the consistency of recall across a workload. 

\rcheck is complementary to both lines of work. 
Rather than modifying embeddings or enforcing fairness at ranking time, \rcheck focuses on disparities that arise within the ANNS process itself: even with fixed embeddings and ranking functions, ANN indexes can produce uneven recall across different regions of the embedding space. 
Moreover, while Wang et al.~\cite{wang2024towards} focus on measuring recall variability, \rcheck detects and reduces it at {\runtime} by adapting search effort across query {\cluster}s. 
We also adopt the proposed robustness as an evaluation metric (i.e., fraction below Ref\_Tgt in \S\ref{sec:eval}) to quantify how effectively \rcheck reduces recall variability. 

\section{Conclusion}
\label{sec:conclusion}

We showed that the modern practice of optimizing for mean recall masks significant differences in recall across queries even when target recall is met, hurting user experience.
To this end, we designed \rcheck, a light-weight \runtime system that identifies low-recall queries and reduces recall differences while achieving high throughput.
\rcheck tunes available search effort parameters, making it readily deployable. 
We evaluated \rcheck using widely-used pgvector~\cite{pgvector}.
At the same throughput, \rcheck reduces recall differences, increasing mean recall by up to \maxRecallatIsoThroughput\%, enabling \maxQueriesMeetingTarget\% more queries to meet target recall compared to the state-of-the-art globally-tuned search configuration.

\bibliographystyle{ACM-Reference-Format}
\bibliography{refs}

\end{document}